\documentclass[preprintnumbers,amsmath,amssymb,preprint]{revtex4}
\usepackage{amssymb}
\usepackage{amsmath}
\usepackage{graphics}
\usepackage{graphicx}
\usepackage{subfigure}
\usepackage{dcolumn}
\usepackage{bm}
\usepackage{hyperref}
\usepackage{enumerate}
\setcitestyle{authoryear,round}

\begin{document}

\preprint{APS/123-QED}

\title{Wall Adhesion and Constitutive Modelling of Strong Colloidal Gels}

\author{D. R. Lester}
\affiliation{CSIRO Mathematics, Informatics and Statistics, PO Box 56, Highett,  Vic. 3190, Australia}
\email{daniel.lester@csiro.au}
\author{R. Buscall}
\affiliation{MSACT Research and Consulting, Exeter, United Kingdom}
\affiliation{Particulate Fluids Processing Centre, Dept. of Chemical and Biomolecular Engineering, University of Melbourne, Victoria 3010, Australia}
\author{A. D. Stickland}
\affiliation{Particulate Fluids Processing Centre, Dept. of Chemical and Biomolecular Engineering, University of Melbourne, Victoria 3010, Australia}
\author{P. J. Scales}
\affiliation{Particulate Fluids Processing Centre, Dept. of Chemical and Biomolecular Engineering, University of Melbourne, Victoria 3010, Australia}

\date{\today}

\begin{abstract}
Wall adhesion effects during batch sedimentation of strongly flocculated colloidal gels are commonly assumed to be negligible. In this study \emph{in-situ} measurements of gel rheology and solids volume fraction distribution suggest the contrary, where significant wall adhesion effects are observed in a 110mm diameter settling column. We develop and validate a mathematical model for the equilibrium stress state in the presence of wall adhesion under both viscoplastic and viscoelastic constitutive models. These formulations highlight fundamental issues regarding the constitutive modeling of colloidal gels, specifically the relative utility and validity of viscoplastic and viscoelastic rheological models under arbitrary tensorial loadings. The developed model is validated against experimental data, which points toward a novel method to estimate the shear and compressive yield strength of strongly flocculated colloidal gels from a series of equilibrium solids volume fraction profiles over various column widths.
\end{abstract}


\maketitle

\section{Introduction}
\label{sec:intro}

The batch settling test is widely utilized as a means to characterize both the sedimentation and consolidation properties of colloidal suspensions~\citep{Kynch:52,Michaels/Bolger:62,Tiller/Shirarto:64,HowellsEA:90,Landman/White:94,Burger/Tory:00,LesterEA:05,Diehl:07,Grassia:11}, where the relevant material properties act as inputs for the modeling of a wide range of solid-liquid separation processes, ranging from tailings disposal and gravity settling, through to continuous thickening and pressure filtration. In the minerals industry, as is exemplified herein, these suspensions are typified by a broad range of particle size distributions, ranging from 0.1 to 200 micron.  To improve the rate of sedimentation, flocculation of the suspensions is augmented through the addition of very high molecular weight ($>$10 million) polymeric flocculants, whereas in other applications, electrolyte coagulants are used to aid aggregation. In either case, the suspensions under consideration are strongly flocculated (often with effective well-depths $>$ 20 kT) and so are non-Brownian (athermal) and stable over long time scales. Hence the behaviour of strongly flocculated suspensions can be very different to that of weak- or partially-aggregated systems; time-dependent phenomena such as spontaneous creep, ripening and collapse cannot occur. As strong systems are simpler, these materials form a benchmark against which more complex, labile weak systems can be compared.

At a critical solids concentration, sometimes as low as a few volume \%, particulate aggregates in strongly flocculated colloidal suspensions can form a continuous space-filling particulate network (or colloidal gel) which can both withstand and transmit stress. This particulate network can consolidate significantly under differential stress (e.g. pressure filtration, gravitation or centrifugation), however as the network is strongly volume-strain hardening, the system reaches an equilibrium concentration (typically significantly less than the close-packing or frictional limit) for a given load.  Conversely, the particulate network is strongly strain-softening in shear, and by the standards of polymer rheology, these systems are very brittle, being able to only withstand shear strains of less than 1\% and quite often less than 0.001\% prior to yield and flow.  Despite the brittle nature of these strongly flocculated suspensions in shear, imposition of shear is not common in traditional batch sedimentation tests.

To characterize the compressive strength of the particulate network, the batch settling test has significant advantages in that it is simple, cheap and highly portable, and the range of compressive stress (typically $\lesssim 1$ kPa) involved is commensurate with many gravity settling applications. The most commonly measured data is the height of the transient sediment/supernatant interface over one or more experiments, and in more sophisticated experiments the equilibrium and/or transient local average solids volume fraction profile $\phi$ is also determined via e.g. gamma ray~\citep{LabbettEA:06} or ultrasonic attenuation~\citep{AuzeraisEA:90}. Although deconvolution of the measured data set into accurate estimates of the relevant material properties is not trivial, significant advances~\citep{Grassia:11,LesterEA:05} have been made in recent years regarding this problem, facilitating accurate and complete suspension characterization from a small number of batch settling tests.

An important assumption underpinning these deconvolution techniques (and sedimentation theory in general) is that effects arising from adhesion between the settling suspension and the container wall are negligible, effectively allowing the sedimentation and consolidation processes to be quantified via a one-dimensional vertical force balance. However, as strongly flocculated gels are both strongly cohesive and adhesive, these materials readily adhere to container walls with a wall adhesive shear strength $\tau_w(\phi)$, as is well-known from studies of wall-slip in colloidal suspension rheometry. The assumption of negligible wall adhesion effect is motivated by estimates that $\tau_w(\phi)$ is small in comparison to the suspension compressive yield strength $P_y(\phi)$, both of which serve as inputs for an equilibrium momentum balance~\citep{Michaels/Bolger:62} over the particulate phase in the vertical direction
\begin{equation}
\frac{d P_y}{d\phi}\frac{\partial\phi}{\partial z}-\Delta\rho g\phi+\frac{2\tau_w(\phi)}{R}=0,\label{eqn:1Dbalance}
\end{equation}
where $z$ is the vertical bed depth (downwards from the suspension/supernatant interface), $\Delta\rho$ the interphase density difference, $g$ gravitational acceleration constant, and $R$ is the radius of the settling container. As the apparent wall adhesion strength $\tau_w(\phi)$ typically appears~\citep{SethEA:08,BuscallEA:93,Barnes:95} to be somewhat smaller but of the same order to the bulk suspension shear yield strength $\tau_y(\phi)$, and the ratio of shear to compressive yield strength $S(\phi)=\tau_y(\phi)/P_y(\phi)$ appears to vary over the range 0.001-0.2~\citep{BuscallEA:87,BuscallEA:88,deKretserEA:02,ZhouEA:01,Channell/Zukoski:97}, these effects may be neglected for all but narrow settling columns.

\begin{centering}
\begin{figure}
\centering
\includegraphics[width=0.45\columnwidth]{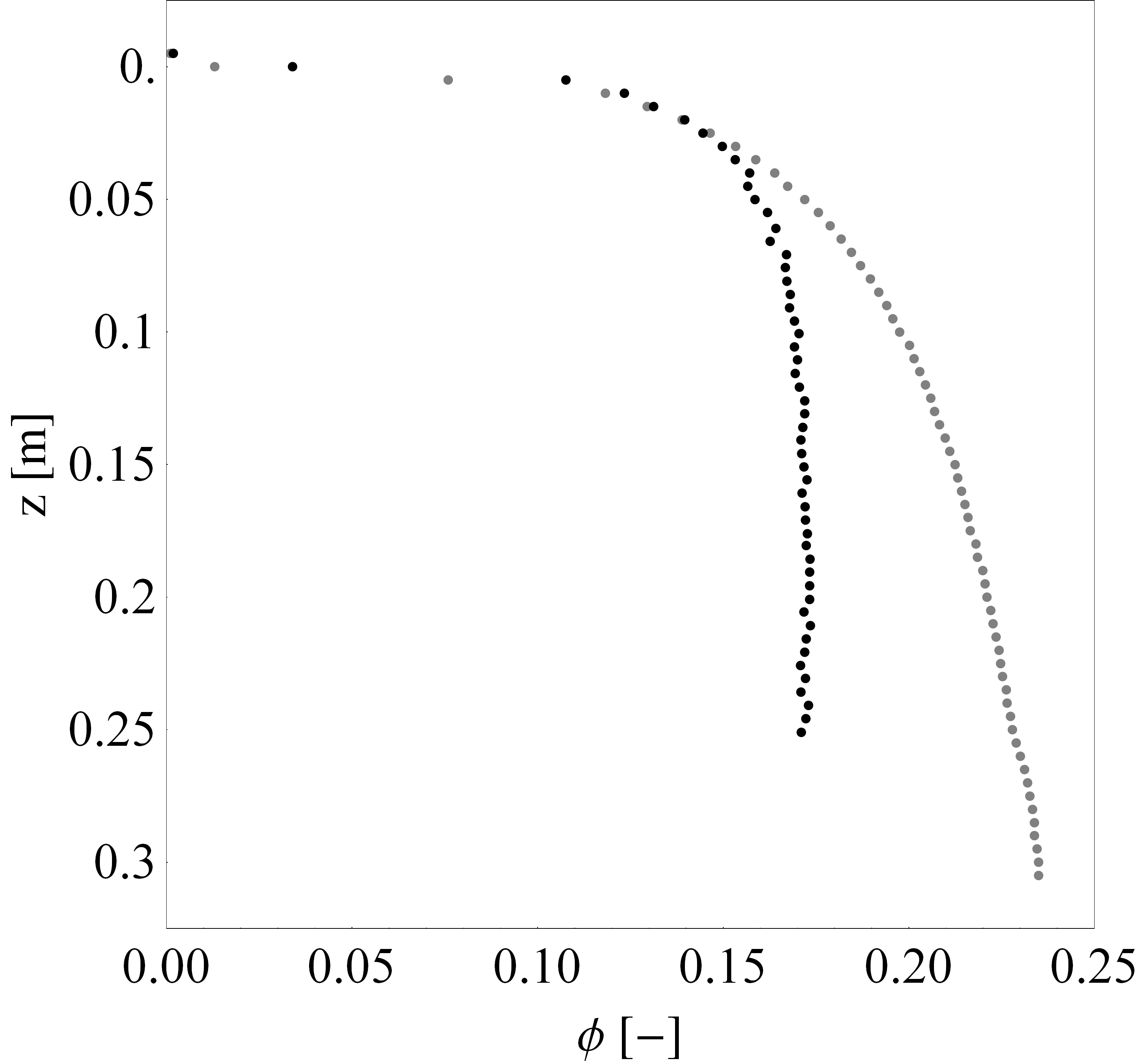}
\caption{Equilibrium solids volume fraction profiles $\phi_\infty$ of a strongly flocculated colloidal gel in narrow ($R_s$=0.011 m, black) and wide ($R_l$=0.055 m, grey) settling columns. Note that a constant solids volume fraction profile can only be explained by wall adhesion supporting the particulate phase.}\label{fig:profiles}
\end{figure}
\end{centering}

However, if wall adhesion effects are significant - i.e. if $\tau_w$ is large and/or $R$ is small - then the assumption of a one-dimensional force balance governing the suspension mechanics breaks down. Now the particulate network experiences a combination of both shear and compressive stress, and this arbitrary stress state varies both vertically and radially. From (\ref{eqn:1Dbalance}), it is clear that the wall adhesion strength acts to counteract the gravitational force, and in some cases, the entire suspension weight can be supported by shear stress alone. Under the approximation $\tau_w\approx\tau_y(\phi)$, this state is given in terms of a critical solids concentration $\phi_c$ which only depends upon the container radius and shear yield strength as
\begin{equation}
\tau_y(\phi_c)\approx\frac{1}{2}\Delta\rho g \phi_c R.\label{eqn:phicritical}
\end{equation}
In principle, once the critical volume fraction $\phi=\phi_c$ is reached (at the critical bed depth $z_c$), the network pressure is constant for bed depths beyond $z_c$. Hence, a clear signature of significant wall adhesion effects is given by a constant equilibrium solids volume fraction profile. Such behaviour is clearly illustrated in Fig.~\ref{fig:profiles}, which depicts the vertical solids volume fraction profiles of a polymer flocculated calcium carbonate suspension in 22 mm and 110 mm diameter columns - the vertical profile for the 22 mm column can only be reasonably explained by wall adhesion effects. As the shear yield strength is a nonlinear monotonic increasing function of $\phi$, this critical state $\phi=\phi_c$ is reached by strongly adhesive colloidal suspensions in narrow containers, given sufficient bed depth. As this critical state is approached ($\phi\rightarrow\phi_c$), estimates of the compressive yield strength $P_y(\phi)$ which neglect wall adhesion effects diverge to $+\infty$, hence wall adhesion can introduce unbounded errors in estimates of suspension material parameters. Such errors can also contaminate the estimate of other suspension properties such as the hindered settling function $R(\phi)$~\citep{LesterEA:05} which quantifies the hydrodynamic drag between particulate and fluid phases.

The data presented herein on the gravity batch settling of mineral particles flocculated with high molecular weight polymers have suggested that wall adhesion effects are by no means always secondary or insignificant. Here the suspensions are seldom far from the gel-point and the ratio of shear to compressive strength is expected to be at a maximum of order unity at the gel-point, decreasing rapidly away from it~\citep{Buscall:09}. Note that for strongly-flocculated colloidal suspensions the wall adhesion force arises directly from the ``sticky'' nature of the flocculated particles, and for the range of stresses typical of batch settling applications, frictional forces do not contribute to wall adhesion. Wall adhesion (typically weaker than particle cohesion) also arises as wall slip in shear rheometry of colloidal suspensions; ironically the lack of total adhesion is problematic in shear rheology, whilst the presence of adhesion is problematic in compressive rheology (i.e. batch settling).

In-situ measurements (i.e. within the settling container itself) of the shear yield strength $\tau_y$ of flocculated colloidal suspensions appear to be significantly higher than those for a decanted suspension. Colloidal gels flocculated with high molecular weight polymer flocculants exhibit rapid irreversible breakdown under shear, and so significant degradation can occur during the decanting process. Conversely, the compressive yield stress is typically measured \emph{in-situ}, hence such inconsistency can significantly underestimate the magnitude of $S(\phi)$. The results herein demonstrate wall adhesion effects are significant in a 110 mm diameter settling column, typically considered to be wide enough to render such effects negligible.


These observations suggest that wall adhesion effects for colloidal gels in batch settling tests are more prevalent than previously appreciated, and have motivated us to investigate the problem of wall adhesion in batch sedimentation in greater detail. In particular, we aim to develop and validate a mathematical model of the suspension equilibrium stress state in the presence of wall adhesion, and develop error estimates for the one-dimensional approximation (\ref{eqn:1Dbalance}) under such conditions. Analysis of the governing multidimensional force balance and suspension behaviour under arbitrary tensorial loadings also raises fundamental questions regarding the constitutive modeling of strongly flocculated colloidal suspensions, particularly the validity and utility of viscoplastic rheological models as opposed to more general but less mathematically tractable viscoelastic formulations. Strongly flocculated colloidal gels exhibit a wide array of complex rheological behaviour~\citep{GrenardEA:14,SprakelEA:11,LindstomEA:12,GibaudEA:10,Santos:13,Koumakis:11,GibaudEA:08,Ovarlez:13,RamosEA:01,Ovarlez:07,CloitreEA:00,Tindley:07,Kumar:12,Uhlherr:05}, including nonlinear creep and time-dependent yield under small shear strains, followed by rapid strain-softening which is described as shear yield prior to viscous flow. Whilst constitutive modelling is still being developed to resolve such complex flow phenomena, the different constitutive approaches (broadly categorized as viscoplastic and viscoelastic models) constitute different levels of resolution of the rheology of colloidal gels.

In this study we find such issues are also of direct relevance with respect to resolution of the wall adhesion problem, hence we consider the properties of each constitutive framework, and utilize an appropriate combination to resolve the wall adhesion problem. Specifically, we seek a macroscopic phenomenological description of the suspension rheology which quantifies the wall adhesion problem in the simplest manner possible, and make no claim as to the network behaviour at the particle level. As such, we seek a minimum extension of traditional 1D compressive rheology of strongly flocculated colloidal suspensions which is capable of addressing the wall adhesion problem. The nature of traditional critical state compressive rheology and the tensorial nature of the wall adhesion problem means that a large-strain visco-elastic framework is required to resolve this problem. This solution is then compared with independent experimental measurements, validating the constitutive approach used herein, and pointing to a novel method to extract accurate estimates of both the compressive $P_y(\phi)$ and shear $\tau_y(\phi)$ yield strength from a series of batch sedimentation tests in columns of varying diameter. Whilst the static equilibrium problem of wall adhesion in batch settling appears to be somewhat divorced from dynamic suspension rheology, these fundamental issues with regard to constitutive modelling apply more broadly to colloidal suspension rheology in general.

In the following Section we develop governing equations for the wall adhesion problem and review constitutive modeling approaches for strongly flocculated colloidal gels. In Sections \ref{sec:hyperelastic} and \ref{sec:viscoplastic} the hyper-elastic and viscoplastic constitutive models respectively are examined in greater detail, and in Section \ref{sec:closure} we present a closure approximation and solution of the viscoplastic model for the equilibrium stress state. A small strain solution of the hyperelastic model is presented by relaxation of the viscoplastic closure in Section \ref{sec:small}, and in Section \ref{sec:application} we validate the viscoplastic solution against experimental data, before conclusions are made in Section \ref{sec:conclusions}.

\section{Constitutive Modeling of Colloidal Gels}
\label{sec:constit}

To develop a quantitative model of batch settling in the presence of wall adhesion effects, we consider the transient dynamics of an attractive colloidal gel within a batch settling experiment starting at the initial condition $\phi=\phi_0<\phi_g$. Over the past few decades, several phenomenological theories of the behaviour of strongly flocculated colloidal gels have been developed~\citep{Richardson/Zaki:54,KimEA:07,Philip:82,Toorman:96,HowellsEA:90,Burger/Concha:98,AuzeraisEA:90,Buscall/White:87} across a variety of diverse fields, with a significant degree of duplication and fragmentation. The majority of these formulations have focussed upon a one-dimensional force balance between the solid and fluid phases, and whilst these have been very successful in capturing the gross features of 1D processes such as pressure filtration and continuous thickening, their ability to resolve multi-dimensional phenomena such as the wall adhesion problem is limited.

Despite these limitations, these constitutive models utilize several simplifying assumptions which provide significant insights into the nature of colloidal suspensions under arbitrary tensorial loads. The most significant of these assumptions is that the impact of anisotropy at the particle scale due to consolidation history is negligible with respect to macroscopic rheology. Whilst consolidation processes such as 1D pressure filtration in confined domains involve uniaxial rather than isotropic consolidation, microstructural re-arrangement via collapse and buckling of particle chains acts to maintain isotropy of the particulate network ~\citep{SetoEA:13}. This assumption, central to compressive rheology, is further supported by the fact that strongly flocculated colloidal gels can only support small deviatoric strains prior to yield, and so the macroscopic rheology is essentially identical under confined uniaxial compression and isotropic volumetric strain, quantified in terms of the solids volume fraction $\phi$.

A multi-dimensional theory of the flow and separation of flocculated colloidal suspensions has been developed~\citep{LesterEA:10} which quantifies the evolution of $\phi$ as
\begin{equation}
\frac{\partial\phi}{\partial t}+\mathbf{q}\cdot\nabla\phi=\nabla\cdot\frac{(1-\phi)^2}{R(\phi)}\left(\Delta\rho\phi\frac{D_\mathbf{q}\mathbf{q}}{Dt}-\nabla\cdot\Sigma^N-\Delta\rho\mathbf{g}\phi\right),\label{eqn:evoln}
\end{equation}
where $D_{\mathbf{q}}/Dt$ is the material derivative with respect to the volume-averaged suspension velocity $\mathbf{q}$, $R(\phi)$ is the hindered settling function or (or inverse Darcy permeability)~\citep{HowellsEA:90,Burger/Concha:98,Buscall/White:87} which quantifies interphase drag, and $\Sigma^N$ is the network stress tensor, defined~\citep{Batchelor:77} as the difference between the total suspension stress $\Sigma$ and the fluid stress $\Sigma^f$:
\begin{equation}
\Sigma^N\equiv\Sigma-\Sigma^f,\label{eqn:networkstress}
\end{equation}
which may be decomposed in terms of the network pressure $p_N$ and deviatoric stress $\boldsymbol\sigma_N$
\begin{equation}
\Sigma^N=-p_N\mathbf{I}+\boldsymbol\sigma_N.
\end{equation}
Under the assumption that during sedimentation the bulk suspension velocity $\mathbf{q}$ is zero, the network force balance simplifies to
\begin{equation}
\frac{\partial\phi}{\partial t}+\nabla\cdot\frac{(1-\phi)^2}{R(\phi)}\left(\nabla\cdot\Sigma^N+\Delta\rho\mathbf{g}\phi\right)=0,\label{eqn:simpleevoln}
\end{equation}
which at equilibrium yields a balance between the network stress gradient and gravitational force
\begin{equation}
\nabla\cdot\Sigma^N+\Delta\rho\mathbf{g}\phi=0.\label{eqn:eqmbalance}
\end{equation}
Central to this model is the specification of a constitutive equation for the colloidal suspension network stress tensor $\Sigma^N$ to close the transient (\ref{eqn:evoln}) and equilibrium (\ref{eqn:eqmbalance}) momentum balances. Henceforth we explore several constitutive modeling approaches for strongly flocculated colloidal gels.

As mentioned in the Introduction, particulate aggregates in colloidal suspensions form a particulate network (or colloidal gel) with finite strength above a critical solids concentration termed the gel point $\phi_g$. For strongly flocculated suspensions such as coagulated or polymer flocculated suspensions, the network stress tensor $\Sigma^N$ is identically zero for $\phi<\phi_g$, whereas for $\phi>\phi_g$, attractive inter-particle forces result in an apparent network strength (in shear and/or differential compression) which strongly increases with solids volume fraction. Hence suspension sedimentation (which involves the settling of hydrodynamically interacting particulate aggregates) occurs for $\phi<\phi_g$, whereas suspension consolidation (which involves simultaneous compression of the particulate network and hydrodynamic drainage) occurs in the range $\phi\geqslant\phi_g$. The athermal nature of the continuous particulate network imparts solid-like properties to strongly flocculated colloidal gels, which leads to a rich array of complex rheological behaviour~\citep{deKretserEA:02,ZhouEA:01,Channell/Zukoski:97,Tindley:07,Kumar:12,Uhlherr:05,GrenardEA:14} under both shear and compressive loads.

Of the array of constitutive models for the network stress $\Sigma^N$ of colloidal gels, there exist two distinct modelling approaches which are most clearly delineated via description of network compression. Flocculated colloidal gels are strongly strain-hardening in compression due to the increase in local solids volume fraction, and so may be described as poro-elastic materials with a volumetric strain-hardening compressional bulk modulus $K(\phi)$. As the inter-particle potential of a strongly flocculated colloidal gel typically contains a deep energy well, the compression of such gels is essentially irreversible. Some workers~\citep{Buscall:09,LietorEA:09,ManleyEA:05,KimEA:07} describe such gels as ``ratchet poro-elastic'', which quantifies the evolution of the network pressure $p_N$ as
\begin{equation}
p_N=\int_{\phi_0}^{\phi}K(\varphi)d\ln\varphi,\,\,\frac{D_s\phi}{Dt}\geqslant 0,\label{eqn:Kcompress}
\end{equation}
where $\phi_0$ is the initial concentration, and $D_s/Dt$ denotes the material derivative with respect to the particulate phase.

Alternately, compressional behaviour of colloidal gels is described by several workers~\citep{Philip:82,Toorman:96,HowellsEA:90,Burger/Concha:98,AuzeraisEA:90,Buscall/White:87} as a viscoplastic process in terms of the so-called compressive yield strength $P_y(\phi)$, which implicitly encodes the irreversible nature of compression. The terminology ``compressive yield'' is somewhat misleading in that it implies an elastic strain limit, whereas in reality the particulate network strain-hardens without limit, and so $P_y(\phi)$ represents the volumetric strain (given by $\phi$) at which an applied network pressure is in equilibrium with the strength of the particulate network:
\begin{equation}
p_N=P_y(\phi),\,\,\frac{D_s\phi}{Dt}\geqslant 0.\label{eqn:Pcompress}
\end{equation}
Hence, in terms of compressive strength, the poro-elastic and viscoplastic formulations are equivalent under the approximation
\begin{equation}
P_y(\phi)\approx P(\phi,\phi_0)\equiv\int_{\phi_0}^{\phi}K(\varphi)d\ln\varphi,\label{eqn:compress_approx}
\end{equation}
which is exact for $\phi_0<\phi_g$. For $\phi_0>\phi_g$, the compressive yield strength is not a true material property, but rather an experimental artefact as it is not dependent upon the initial volume fraction $\phi_0$, i.e. $P_y\neq P$. However, for $\phi_0-\phi_g\ll 1$, the compressive yield strength represents an accurate approximation to the true network pressure $P$. This approximation typifies the relationship between the poro-elastic and viscoplastic formulations; whilst the former more accurately reflects the colloidal gel rheology, the latter leads to more tractable formulation as the strain history need not be evaluated.

In contrast to compression, the shear response of particulate gels is strain-softening, typically, and hence not self-limiting. Indeed, many colloidal gels strain-soften so rapidly that they can be considered to yield. Where this is the case, the notion of a critical yield stress $\tau_y$, or, sometimes, a yield strain $\gamma_c$, is adequate for many purposes. While detailed experiments~\citep{Uhlherr:05,Tindley:07,Kumar:12} indicate particulate gels actually undergo this transition over a range of stresses and strains, suggesting the true yield criterion is more complicated than a critical stress or strain condition, one can still identify a representative critical strain $\gamma_c$ associated with the rapid transition to viscous flow. Colloidal gels are typically brittle in shear, and the representative critical strain is of the order $10^{-4}-10^{-2}$~\citep{Channell/Zukoski:97,BuscallEA:87,Uhlherr:05,Tindley:07}.

In many applications (including experimental studies and numerical simulations) it is often neither feasible nor desirable to resolve such strains and the detailed sub-yield dynamics of colloidal gels. In this case, a viscoplastic constitutive model (such as a Herschel-Bulkley or Bingham model) serves as a useful engineering approximation for the deviatoric network stress tensor $\boldsymbol\sigma_N$, which for simple shear may be quantified as
\begin{equation}
\tau_N=\left(\frac{\tau_y(\phi)}{\dot\gamma}+\eta(\phi,\dot\gamma)\right)\dot\gamma\,\,\,\text{for}\,\,\tau_N\geqslant\tau_y(\phi),\label{eqn:taushear}
\end{equation}
where $\tau_N$ is the 2nd invariant of $\boldsymbol\sigma_N$, $\eta(\phi,\dot\gamma)$ is the apparent suspension viscosity (which is typically non-Newtonian), and $\dot\gamma$ is the rate of shear strain. One disadvantage of the viscoplastic constitutive model is that in general the deviatoric stress $\boldsymbol\sigma_N$ is unresolved below the yield stress $\tau_N<\tau_y(\phi)$, as $\dot\gamma\rightarrow 0$ and the effective viscosity diverges. Although specialized regularization methods have been developed for numerical calculations~\citep{Balmforth:2014}, the conceptual problem of an undefined sub-yield stress state persists.

Conversely, the poro-elastic constitutive model resolves the detailed elastic strain in terms of the shear modulus $G(\phi,\gamma)$ and memory function $m(t)$ via the quasi-linear viscoelastic~\citep{Fung:93} constitutive model
\begin{equation}
\tau_N=\int_{-\infty}^t m(t-s)G(\phi,\gamma)\frac{\partial\gamma}{\partial s}ds+\eta(\phi,\gamma)\dot\gamma\,
\label{eqn:Gshear}
\end{equation}
in which time-strain separability has been invoked as a first approximation for the sake of clarity; most real colloidal gets are not expected to be so obliging, necessarily, even though there are examples, remarkably~\citep{Yin_Solomon:08}. Strain softening is encoded via the shear modulus $G(\phi,\gamma)$, and the strain rate $\dot\gamma$ is small prior to strain softening, which is interpreted as yield in the viscoplastic model. Hence for rapid shear strain (i.e. significantly faster than the relaxation timescale of $m(t)$), the shear yield stress and shear modulus are related via the critical strain as
\begin{equation}
\tau_y(\phi)=\int_0^{\gamma_c} G(\phi,\gamma)d\gamma.\label{eqn:tau_G}
\end{equation}

As such, the viscoplastic and poro-elastic constitutive models can be reconciled as different levels of approximation for the rheology of a colloidal gel, with distinct advantages and disadvantages. Whilst the tensorial form of the poro-elastic model (detailed in Section 3) is a more accurate representation of the dynamics of a strongly flocculated colloidal gel, the viscoplastic model represents a lower-order approximation which has utility in a wide range of applications.

In this study we are primarily interested in solution of the equilibrium stress state, however under the poro-elastic formulation the transient problem (\ref{eqn:simpleevoln}) must be evolved from the initial condition ($\phi=\phi_0$) toward the limit $t\rightarrow\infty$ to determine the distribution of stress and strain at the equilibrium state. Furthermore, as the strains associated with consolidation are large, finite strain measures are required to track material displacements in the Lagrangian frame, which adds further computational complexity.
Conversely, the viscoplastic formulation allows one to analyse the equilibrium state (\ref{eqn:eqmbalance}) directly without need for temporal evolution. However, in multiple dimensions the viscoplastic model can lead to an under-determined stress state, analogous to statically indeterminate problems in structural mechanics. To circumvent this problem, we use a combination of both formulations to address the wall adhesion problem, which greatly simplifies the solution methodology.

\section{Hyperelastic Constitutive Model}
\label{sec:hyperelastic}

The poro-elastic constitutive model under 1D compression (\ref{eqn:Kcompress}) or simple shear (\ref{eqn:Gshear}) may be extended to arbitrary tensorial loadings via a hyperelastic constitutive model which is general enough to capture most observed phenomena of colloidal gels~\citep{GrenardEA:14,SprakelEA:11,Uhlherr:05,Kumar:12}. As particulate gels can undergo large volumetric strains, a finite strain measure is required as a basis for the hyperelastic model, as is provided by the Hencky strain tensor $\mathbf{H}=\ln\mathbf{U}$, where $\mathbf{U}$ is the right stretch tensor, i.e. $\mathbf{F}=\mathbf{R}\mathbf{U}$ where $\mathbf{R}$ is a proper orthogonal tensor, and $\mathbf{F}=\frac{\partial\mathbf{x}}{\partial\mathbf{X}}$ is the deformation gradient tensor arising from the Eulerian $\mathbf{x}$ and Lagrangian $\mathbf{X}$ coordinate frames. The Hencky strain tensor provides a convenient basis for constitutive modelling as $\mathbf{H}$ is work-conjugate with the Cauchy stress tensor $\Sigma^N$. Furthermore, the set of modified invariants $K_i$, $i=1:3$ of $\mathbf{H}$ introduced by \cite{Criscione:00} give rise to response terms which are mutually orthogonal, providing a clear elucidation between the invariants and various modes of deformation and their underlying symmetries. The first such invariant $K_1$ is associated with volumetric strain
\begin{equation}
K_1=\text{tr}(\mathbf{H})=\ln\frac{\phi_0}{\phi},\label{eqn:K1}
\end{equation}
whilst the second invariant $K_2$ quantifies the magnitude of shear strain
\begin{equation}
K_2=\sqrt{\text{dev}(\mathbf{H}):\text{dev}(\mathbf{H})},\label{eqn:K2}
\end{equation}
where $\mathbf{H}=\frac{1}{3}K_1\mathbf{I}+K_2\boldsymbol\Phi$, and the normalized deviatoric strain $\boldsymbol\Phi=\text{dev}(\mathbf{H})/K_2$, with $\boldsymbol\Phi:\boldsymbol\Phi=1$. The third invariant $K_3$ is associated with the mode of distortion
\begin{equation}
K_3=3\sqrt{6}\text{det}(\boldsymbol\Phi),\label{eqn:K3}
\end{equation}
where $K_3\in[-1,1]$ such that $K_3=-1$ corresponds to uniaxial extension, $K_3=1$ uniaxial compression, and $K_3=0$ to pure shear.

The original hyperelastic model is based upon an elastic potential $\psi=\psi(K_1,K_2,K_3)$ which for perfectly elastic materials stores all work done by material deformations as internal strain energy. For such materials, the isotropic Cauchy stress $\mathbf{t}$ is
\begin{equation}
 J\mathbf{t}=\frac{\partial\psi}{\partial\mathbf{H}},\label{eqn:Hencky_elastic}
\end{equation}
where $J=\text{det}(\mathbf{F})$ is the total volumetric strain. The hyperelastic framework can also be extended to dissipative materials (as per the K-BKZ or Rivlin-Saywers type viscoelastic models), in which case the potential $\psi$ loses its strict thermodynamic interpretation (via decomposition into conservative and dissipative components $\psi=\psi_c+\psi_d$) as strain energy is no longer fully conserved, leading to irreversible deformations characteristic of strongly flocculated colloidal suspensions.

In general, the potential $\psi$ may be dependent upon both strain-rate and strain history, as per the invariants $K_i$
\begin{equation}
\psi=\psi(K_i,\dot{K}_i,t-s),\label{eqn:psi_general}
\end{equation}
where $s\in(-\infty,t)$ is the strain history. As for a general nonlinear viscoelastic material~\citep{Wineman:09}, the network stress $\Sigma^N$ is given by the generalisation of the Cauchy stress in (\ref{eqn:Hencky_elastic}) for a dissipative materials as
\begin{equation}
\Sigma^N=\mathcal{H}[\psi(K_i,\dot{K}_i,t-s)]_{-\infty}^t.\label{eqn:strain_history}
\end{equation}
where the functional $\mathcal{H}[\,\,]^t_{-\infty}$ acts over the entire strain history. As such, the hyperelastic potential $\psi$ encodes the full viscoelastic rheology of the particulate network, including compressive and shear deformations and tensorial combinations thereof. For this dissipative potential $\psi$, from (\ref{eqn:strain_history}) the network stress tensor is quantified via the integro-differential equation
\begin{equation}
\begin{split}
\Sigma^N=&\int_{-\infty}^t\frac{1}{\exp{K_1}}\sum_{i=1}^3\frac{\partial\psi}{\partial \mathbf{H}}ds\\
=&\int_{-\infty}^t\frac{1}{\exp{K_1}}\left(\frac{\partial\psi}{\partial K_1}\mathbf{I}+\frac{\partial\psi}{\partial K_2}\boldsymbol\Phi-\frac{1}{K_2}\frac{\partial\psi}{\partial K_3}\mathbf{Y}\right)ds,\label{eqn:hyperelastic}
\end{split}
\end{equation}
where $\mathbf{Y}=3\sqrt{6}\boldsymbol\Phi^2=\sqrt{6}\mathbf{I}-3K_3\boldsymbol{\Phi}$. Hence $\partial\psi/\partial K_i$ encode the rheological properties of the suspension, namely the shear and compressive moduli. As strongly flocculated colloidal gels can only support small deviatoric strains prior to yield, then the compressive behaviour is essentially identical under differential uniaxial compression ($K_3=-1$) or spherical volumetric strain ($K_2=0$, $K_3$ indeterminate), and so the dependence of these materials upon the $K_3$ invariant is negligible. This simplification follows directly from the arguments in Section \ref{sec:constit} that the macroscopic rheology can be quantified in terms of the total volumetric strain (encoded as $K_1$ or $\phi$) alone. Hence the rheology of strong colloidal gels only depends upon the magnitude of the deviatoric ($K_2$) and isotropic ($K_1$) strains. Whilst this simplification does not necessarily preclude interaction between combined shear and compression loadings, and experimental evidence~\cite{Channell/Zukoski:97} suggests such interactions can be significant, for simplicity we assume herein that these deformation modes act independently.

Under these assumptions, the isotropic $p_N$ and deviatoric $\boldsymbol\sigma_N$ components of the network stress tensor $\Sigma^N$ may be generalized from (\ref{eqn:Kcompress}) (\ref{eqn:Gshear}) as
\begin{align}
&p_N=\int_{-\infty}^t\frac{1}{\exp K_1}\frac{\partial\psi}{\partial K_1}ds=\int_{-\infty}^{t}\frac{K(\phi)}{\phi}\frac{\partial\phi}{\partial s}ds=\int_{\phi_0}^{\phi(t)}K(\phi)d\ln\phi,\label{eqn:hyperpressure}\\
&\boldsymbol\sigma_N=\int_{-\infty}^t\frac{1}{\exp K_1}\frac{\partial\psi}{\partial K_2}\boldsymbol\Phi ds=\int_{-\infty}^t \frac{\partial G(\phi,\gamma,t-s)}{\partial s}\boldsymbol\Phi(s) ds\label{eqn:hypershear},
\end{align}
where $K(\phi)$, $G(\phi,\gamma,t)$ are the bulk and shear moduli respectively, which are related to $\psi$ as
\begin{align}
&\frac{\partial\psi}{\partial K_1}=-\phi_0\frac{\partial}{\partial s}\left(\frac{K(\phi)}{\phi}\right),\\
&\frac{\partial\psi}{\partial K_2}=\frac{\phi_0}{\phi}\frac{\partial G(\phi,\gamma,t-s)}{\partial s}.
\end{align}

Equations (\ref{eqn:hyperpressure}), (\ref{eqn:hypershear}) represent tensorial forms of (\ref{eqn:Kcompress}), (\ref{eqn:Gshear}) under an appropriate finite strain measure ($\mathbf{H}$) for colloidal gels. This hyperelastic constitutive model describes the solid mechanics of the particulate network as a viscoelastic material which via (\ref{eqn:evoln}) describe the deformation and flow and separation of colloidal gels.


\section{Viscoplastic Constitutive Model}
\label{sec:viscoplastic}

Given appropriate assumptions regarding the evolution of a batch settling experiment under the influence of wall adhesion effects, the viscoplastic constitutive model allows the equilibrium state $\phi=\phi_\infty$ to be approximated directly via the force balance (\ref{eqn:eqmbalance}) without need to solve the full material evolution equation (\ref{eqn:evoln}). The primary assumption underpinning the equilibrium state is that the suspension is in a \emph{critical} state, whereby the network pressure $p_N$ is balanced by the compressive yield strength $P_y(\phi_\infty)$ throughout
\begin{equation}
p_N=P_y(\phi_\infty),\label{eqn:pyphi}
\end{equation}
This critical state arises for all colloidal gels regardless of the reversibility of consolidation,  if \emph{(i)} the suspension is initially unnetworked, i.e. $\phi_0\leqslant\phi_g$, and \emph{(ii)} the network pressure $p_N$ for each material element monotonically increases with time over the course of the experiment. Under these conditions, the network pressure $p_N$ in (\ref{eqn:pyphi}) is identical to that of the hyperelastic formulation (\ref{eqn:hypershear}). This assumption is supported by the fact that the hydrodynamic drag between phases in a batch settling experiment decreases monotonically with time, from the initial condition in an asymptotic fashion toward the equilibrium state when the gravitational stress is supported solely by the inherent strength of the particulate network.

During the sedimentation and consolidation process, fluid upflow at the walls (via a lubrication film) can be observed which prevents the network from adhering. It is proposed that as the equilibrium state is approached, the shear stress associated with this film generates a microscopic stick-slip mechanism between the particulate network and the container wall which allows the suspension interface to subside so long as this shear stress exceeds the wall adhesion strength. This mechanism is supported by experimental observations that the equilibrium suspension/supernantant interface is flat, whereas adhesion without slip would generate a concave interface due to subsidence of material in the interior. As the compressive stress due to gravitational acceleration increases with depth, this flat interface means that the particulate network experiences a compressive load throughout, whereas a convex interface could impart tensile load near the walls. The stick/slip mechanism suggests that the suspension is in a critical state of shear stress at the container walls, i.e. the suspension shear stress is equivalent to the wall adhesion strength
\begin{equation}
\tau_N|_{r=R}=\tau_w(\phi|_{r=R}).\label{eqn:wallcondition}
\end{equation}
This wall adhesion boundary condition for strongly flocculated colloidal suspensions is well-established in shear rheometry for the problem of wall-slip. Whilst it is conceivable that at large network pressures $p_N$, Coulombic friction may augment the wall boundary condition, the compressive stresses typical of batch settling experiments are such that frictional effects are dominated by the strong adhesive force between flocculated particles and the container wall. As this adhesive force (which gives rise to $\tau_w$) is weaker than the cohesive force between particles (which gives rise to $\tau_y$), this wall adhesion boundary condition ensures that throughout the internal shear stress $\tau_N\leqslant\tau_y(\phi)$, and by symmetry the shear stress is zero at the central axis:
\begin{equation}
\tau_N|_{r=0}=0.\label{eqn:axiscondition}
\end{equation}
As the wall adhesion boundary condition ensures that the critical shear strain is never exceeded in the batch settling experiment, the suspension can only undergo very small strains in shear. Conversely, due to consolidation, the suspension can undergo large volumetric strains, as reflected in the change in local solids volume fraction from a few \% to order 20\% at equilibrium. This behaviour is typical of all strongly flocculated colloidal suspensions, these materials are brittle in shear and so can only support small deviatoric strains prior to flow, but are poroelastic in compression and so can support large volumetric strains which are largely irreversible (hence ``ratchet poroelastic'').

Further boundary conditions at the top of the bed are given by
\begin{align}
\phi|_{z=0}=\phi_g,\\
p_N|_{z=0}=0,\\
\boldsymbol\sigma_N|_{z=0}=\mathbf{0}.\label{eqn:topcondition}
\end{align}

For the axis-symmetric batch settling problem, the equilibrium network force balance (\ref{eqn:eqmbalance}) may be directly expanded in cylindrical coordinates $(r,\theta,z)$ as
\begin{align}
&\frac{\partial\Sigma^N_{rr}}{\partial r}+\frac{\Sigma^N_{rr}-\Sigma^N_{\theta\theta}}{r}+\frac{\partial\Sigma^N_{rz}}{\partial z}=0,\label{eqn:force_r}\\
&\frac{\partial\Sigma_{\theta\theta}}{\partial \theta}=0,\label{eqn:force_theta}\\
&\frac{\partial\Sigma^N_{rz}}{\partial r}+\frac{\Sigma^N_{rz}}{r}+\frac{\partial\Sigma^N_{zz}}{\partial z}+\Delta\rho g \phi=0,\label{eqn:force_z}
\end{align}


where $z$ is the vertical coordinate down from the suspension/supernatant interface, hence $\mathbf{g}=g\,\hat{\mathbf{e}}_z$. Due to symmetry the transverse angular stresses $\Sigma^N_{r\theta}$, $\Sigma^N_{z\theta}$ are zero, and all gradients with respect to $\theta$ are zero. We define the network pressure as $p_N=-\frac{1}{3}\text{tr}(\Sigma^N)$, the network shear stress magnitude $\tau_N=|\text{dev}(\Sigma^N)|$, and the first and second normal stress differences respectively are $N_1=\Sigma^N_{zz}-\Sigma^N_{rr}$, $N_2=\Sigma^N_{rr}-\Sigma^N_{\theta\theta}$, following the usual convention where $(z,r,\theta)$ denote the (``flow'', ``gradient'', ``vorticity'') directions for the batch settling problem. As the number of field variables ($\phi$, $\Sigma^N_{rr}$, $\Sigma^N_{\theta\theta}$, $\Sigma^N_{zz}$, $\Sigma^N_{rz}$) exceeds the number of field equations (\ref{eqn:pyphi}), (\ref{eqn:force_r}), (\ref{eqn:force_theta}), (\ref{eqn:force_z}), the system is under-determined.\\

Such statically indeterminate problems are common in plasticity theory; typically these arise from the application of constitutive models which do not possess well-defined stress-strain relationships below the critical yield stress. This deficiency is not a physical problem \emph{per se}, but rather stems from the over-simplified constitutive model; there exists a large class of problems for which plastic models generate under-determined systems~\citep{Hill:50,Balmforth:2014}. A common approach to resolve statically indeterminate systems is to invoke small-strain elasticity to solve deformations away from the equilibrium state and thus determine the equilibrium stress distribution. We utilise a similar approach here in that a closure approximation is invoked to generate a viscoplastic approximation to the equilibrium stress state, which may then be converted into the hyperelastic frame and closure approximation relaxed at the expense of disrupting the force balance away from equilibrium.

As such, the viscoplastic estimate serves as a psuedo-initial condition for the hyperelastic formulation, from which the equilibrium state can be approached via the temporal evolution equation (\ref{eqn:evoln}). If the closure approximation used to generate the viscoplastic estimate is accurate, only small deviatoric strains are required to evolve this estimate toward the true hyperelastic equilibrium condition, greatly simplifying the solution process. Furthermore, propagation toward the true equilibrium state generates quantitative estimates of the accuracy of the viscoplastic approximation.

\section{Closure and Solution of Viscoplastic Formulation}
\label{sec:closure}

A closure approximation for the viscoplastic formulation may be generated by consideration of the equilibrium state below the critical bed depth $z_c$, where the stress equilibrium conditions render all derivatives with respect to $z$ to be zero. Under these conditions the viscoplastic solution simplifies to $\frac{\partial}{\partial r}\Sigma^N_{rr}+(\Sigma^N_{rr}-\Sigma^N_{\theta\theta})/r=0$, and $\frac{\partial}{\partial r}\Sigma^N_{rz}=\Delta\rho g\phi$, which under the axisymmetric boundary condition (\ref{eqn:axiscondition}) gives
\begin{equation}
-p_N-\frac{1}{3}N_1+\frac{1}{3}N_2=\int_0^r \frac{N_2}{r'}dr'-P_y(\phi|_{r=0}),\quad\text{for}\,\,z>z_c.
\end{equation}
As $N_1,N_2\leqslant\tau_N\leqslant\tau_y(\phi)\ll P_y(\phi)$ and the compressive stress $P_y(\phi)$ is a strongly increasing function of $\phi$, the radial solids volume fraction distribution varies weakly as
\begin{equation}
\phi(r)=P_y^{-1}\left[ P_y(\phi|_{r=0})-\int_0^r \frac{N_2}{r'}dr' -\frac{1}{3}N_1+\frac{1}{3}N_2 \right]=\phi|_{r=0}+\delta\phi(r),
\end{equation}
where $|\delta\phi(r)|\ll|\phi_{r=0}|$ for $z>z_c$. Hence the $rz$ stress component $\Sigma^N_{rz}$ also deviates weakly from the linear distribution
\begin{equation}
\Sigma^N_{rz}=-\Delta\rho g\phi|_{r=0}\frac{r^2}{2}+\Delta\rho g\int_{0}^{r}\delta\phi(r')r'dr'.
\end{equation}
These scalings motivate us to consider the approximation $\delta\phi=0$, which corresponds to the assumption that the first and second normal stress differences $N_1$, $N_2$ are negligible throughout the entire suspension, both above and below the critical bed depth $z_c$. Although the validity of this assumption for $z<z_c$ is unknown, this assumption is invoked temporarily as an intermediate step prior to relaxation of this closure under the hyperelastic formulation.

Invoking the closure approximation $N_1=N_2=0$ simplifies the total shear stress to $\tau_N=|\Sigma^N_{rz}|$, and closes the set of governing equations which can be diagonalized in terms of the coupled hyperbolic system
\begin{align}
\frac{\partial\omega_1}{\partial z}+\frac{\partial\omega_1}{\partial r}=-\frac{\omega_1-\omega_2}{2r}+\Delta\rho g f\left(\frac{\omega_1+\omega_2}{2}\right),\label{eqn:omega1}\\
\frac{\partial\omega_2}{\partial z}-\frac{\partial\omega_2}{\partial r}=-\frac{\omega_1-\omega_2}{2r}+\Delta\rho g f\left(\frac{\omega_1+\omega_2}{2}\right),\label{eqn:omega2}
\end{align}
where $\omega_1=p_N+\tau_N$, $\omega_2=p_N-\tau_N$, and $f(p)=P_y^{-1}(p)$. From (\ref{eqn:axiscondition})-(\ref{eqn:topcondition}), the initial and boundary conditions are
\begin{align}
&\omega_1|_{z=0}=\omega_2|_{z=0}=0,\label{eqn:z_zero_bc}\\
&\omega_1|_{r=0}=\omega_2|_{r=0},\label{eqn:r_zero_bc}\\
&\omega_1|_{r=R}=\mathcal{F}\left(\omega_2|_{r=R}\right),\label{eqn:r_R_bc}
\end{align}
where $\mathcal{F}$ describes the relationship between the shear and compressive stress at the container wall, such that $\tau_N=\tau_y(\phi|_{r=R})$, $p_N=P_y(\phi|_{r=R})$. Based upon physical arguments and experimental data, \citet{Buscall:09} proposes a relationship for the ratio $S(\phi)=\tau_y(\phi)/P_y(\phi)$ between the shear and compressive yield strength of a particulate gel, which rapidly decreases from around 1 at the gel point $\phi_g$ to the asymptotic value $S_\infty$ with increasing $\phi$. For a compressive yield strength of the form
\begin{equation}
P_y(\phi)=k\left(\left(\frac{\phi}{\phi_g}\right)^n-1\right),\label{eqn:pyphi_func}
\end{equation}
the asymptotic value is
\begin{equation}
S_\infty=\kappa n \gamma_c.\label{eqn:Sinf}
\end{equation}
where $\gamma_c$ is the critical shear strain and $\kappa$ the ratio of shear to compressive moduli, which is related to the the Poisson ratio $\nu$ as
\begin{equation}
\kappa=\frac{2}{3}\left(\frac{1-\nu}{1-2\nu}\right),
\end{equation}
where $\nu=3/8$, $\kappa=5/3$ for systems bound by central forces as per Cauchy's relationships. For a compressive yield strength with the functional form (\ref{eqn:pyphi_func}), $S(\phi)$ is then
\begin{equation}
S(\phi)=\left(\left(\frac{1}{S_{\infty}}-1\right)\left(1-\left(\frac{\phi}{\phi_g}\right)^{-n}\right)+1\right)^{-1}.\label{eqn:Sphi}
\end{equation}

For these relationships, the operator $\mathcal{F}$ is explicitly
\begin{equation}
\mathcal{F}(\omega_1)=\frac{-S_\infty(2k+\omega_1)+\sqrt{4k S_\infty^2(k+\omega_1)+\omega_1^2}}{1+S_\infty},\label{eqn:F}
\end{equation}
however other forms arise for different functional forms of $P_y(\phi)$.\\

\begin{figure}[h]
\centering
\begin{tabular}{c c}
\includegraphics[width=0.4\columnwidth]{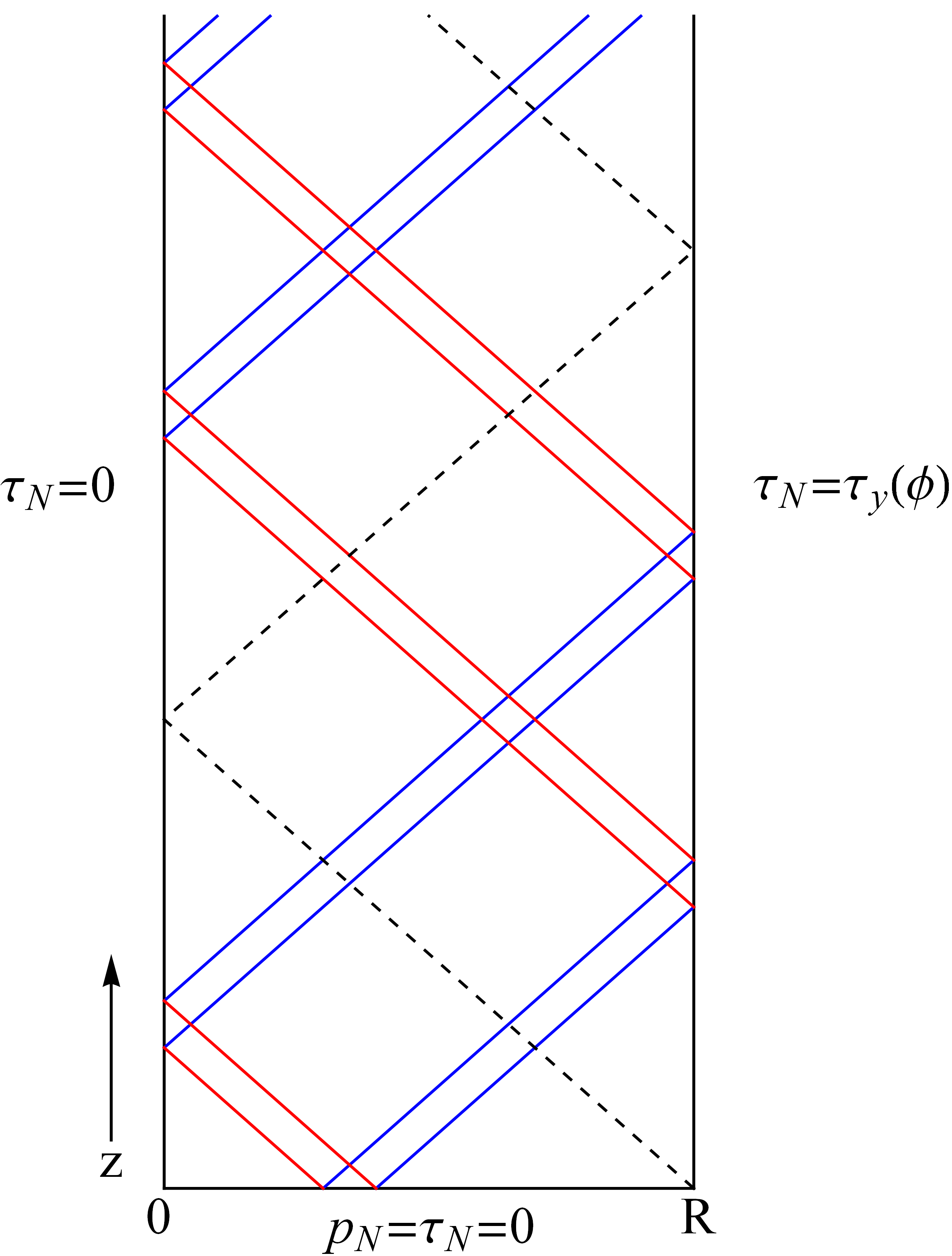}&
\includegraphics[width=0.4\columnwidth]{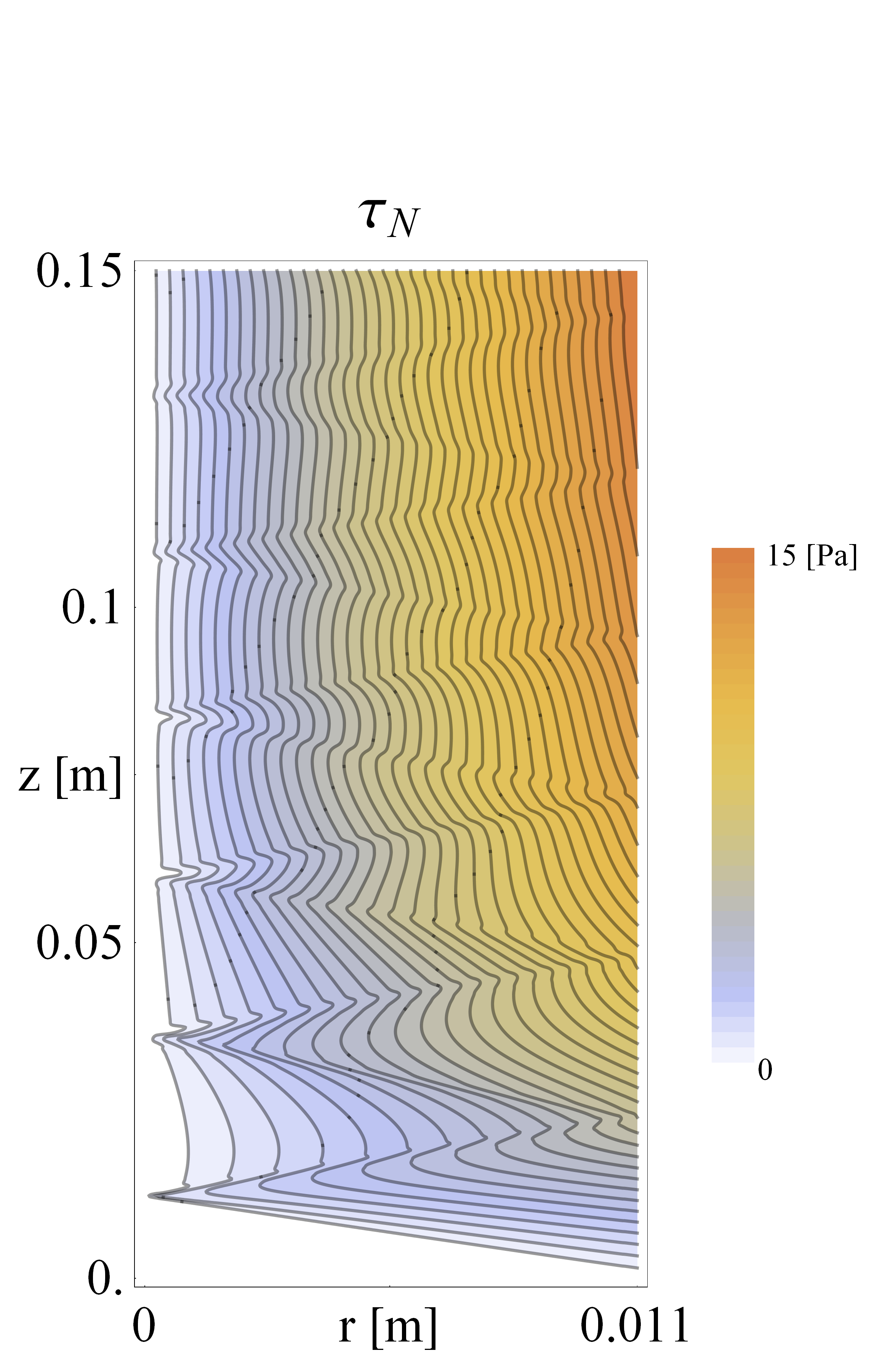}\\
(a) & (b)
\end{tabular}
\caption{Schematic (a) of $\omega_1$ (red), $\omega_2$ (blue) characteristics and boundary conditions for the viscoplastic formulation, and typical distribution (b) of the network shear stress $\tau_N$.}\label{fig:characteristics}
\end{figure}

\begin{figure}
\centering
\begin{tabular}{c c c}
\includegraphics[width=0.3\columnwidth]{tau_magnafloc_small.png}&
\includegraphics[width=0.3\columnwidth]{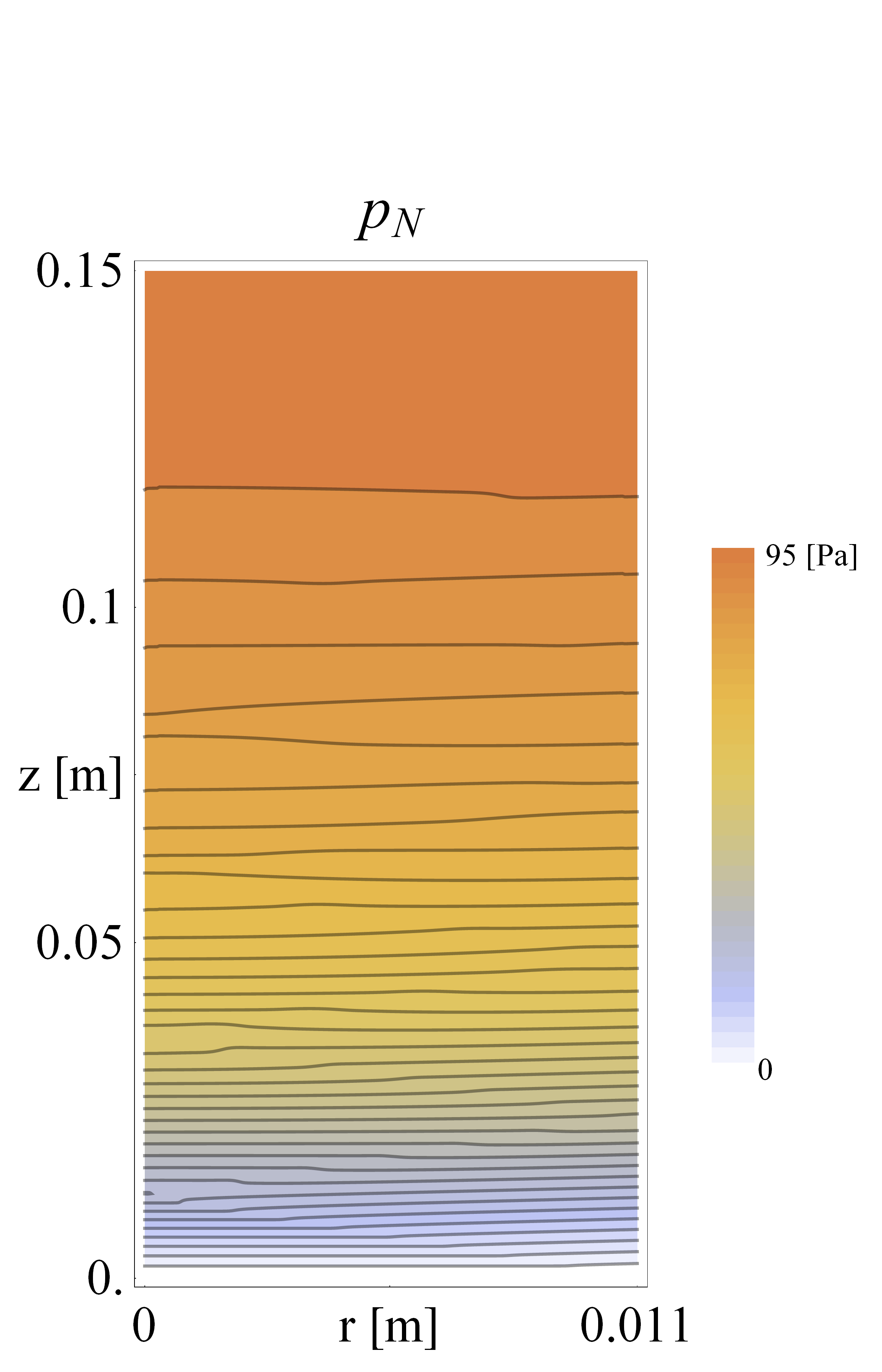}&
\includegraphics[width=0.3\columnwidth]{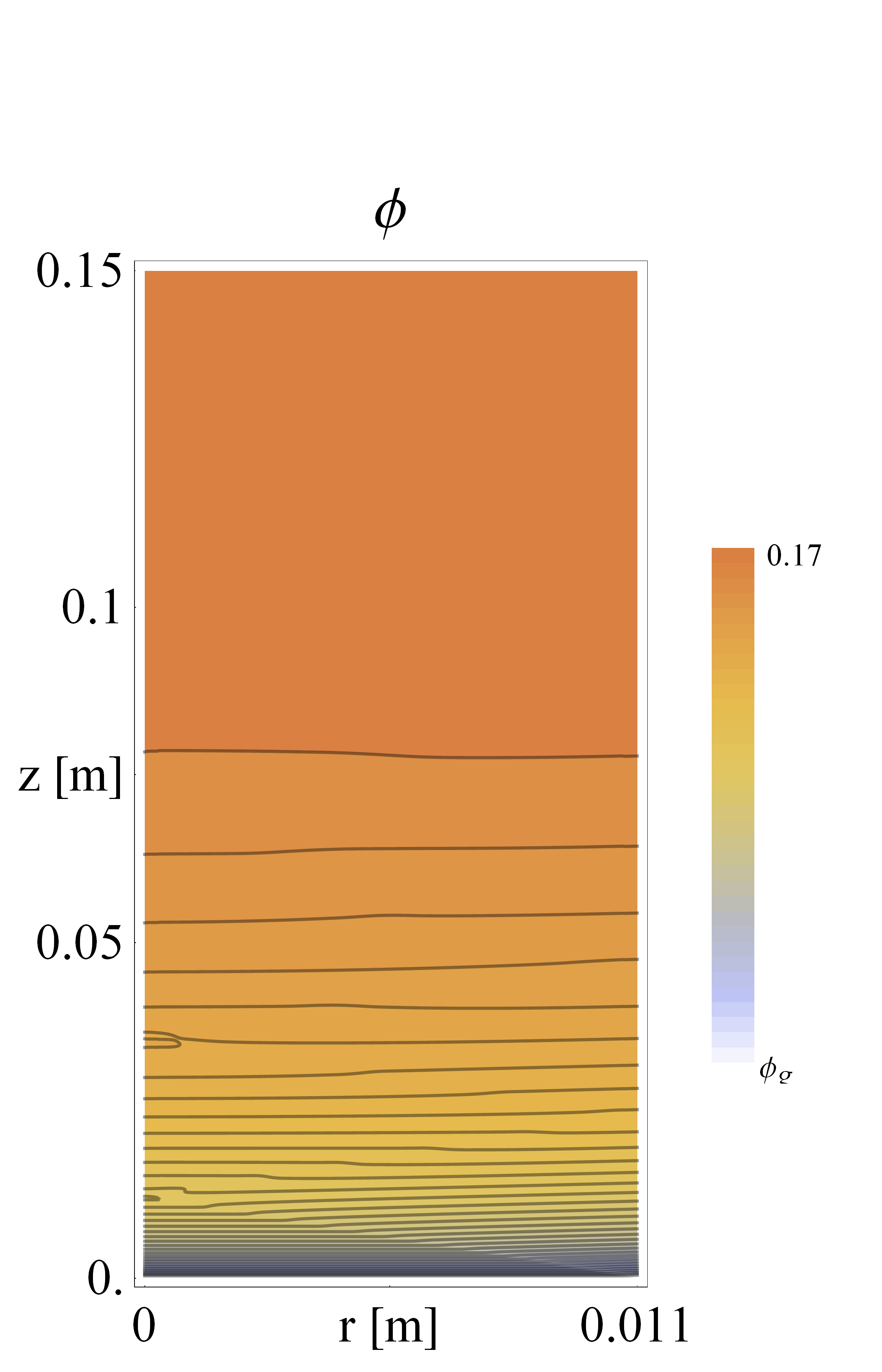}\\
(a) & (b) & (c)
\end{tabular}
\caption{Equilibrium (a) network shear stress $\tau_N$, (b) network pressure $p_N$, and (c) solids volume fraction $\phi$ distributions for suspension (a) in Table~\ref{tab:suspensions} as predicted by viscoplastic formulation in a 22 mm diameter container.}\label{fig:shearpress_small}
\end{figure}

\begin{figure}
\centering
\begin{tabular}{c c c}
\includegraphics[width=0.3\columnwidth]{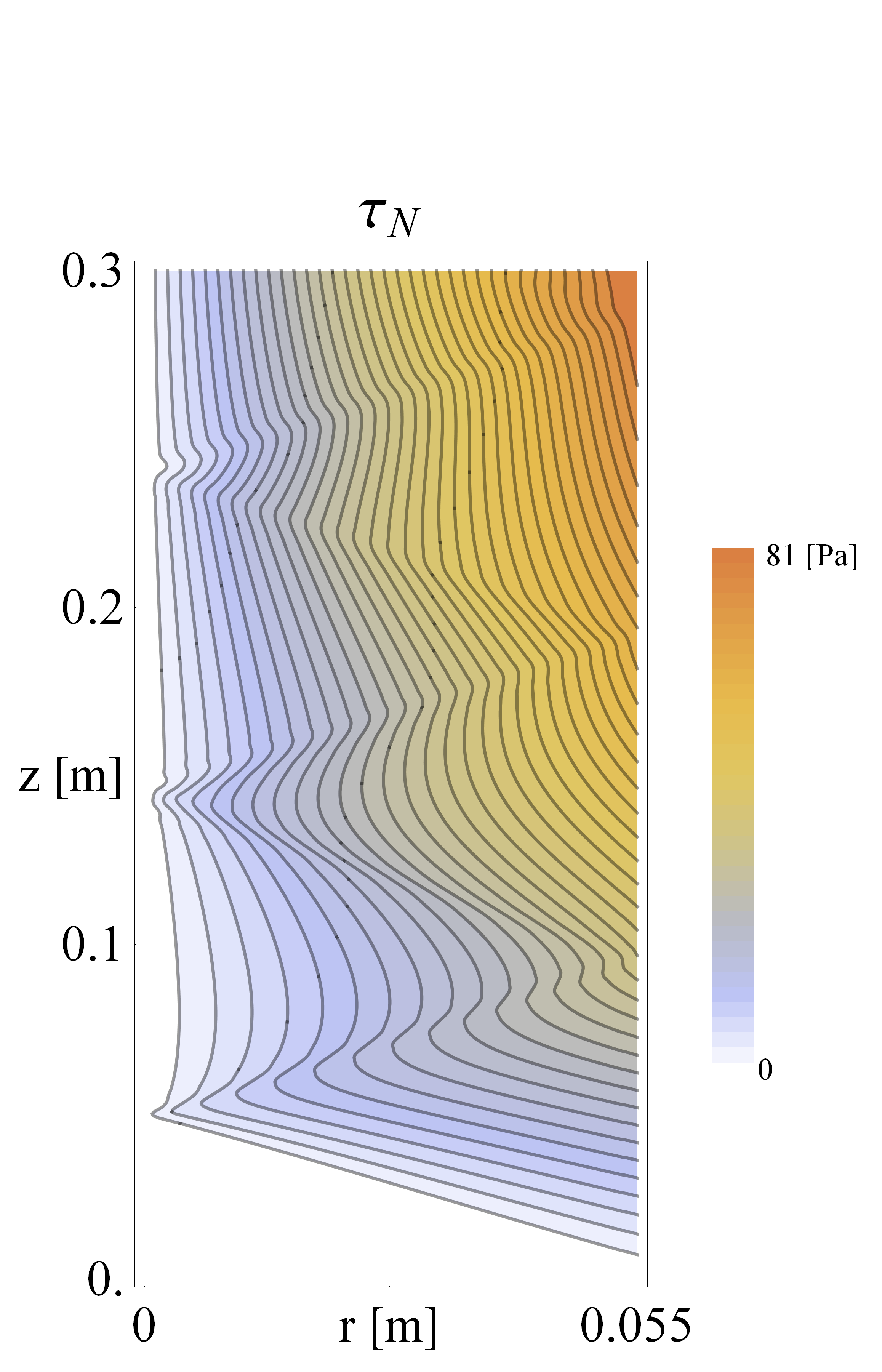}&
\includegraphics[width=0.3\columnwidth]{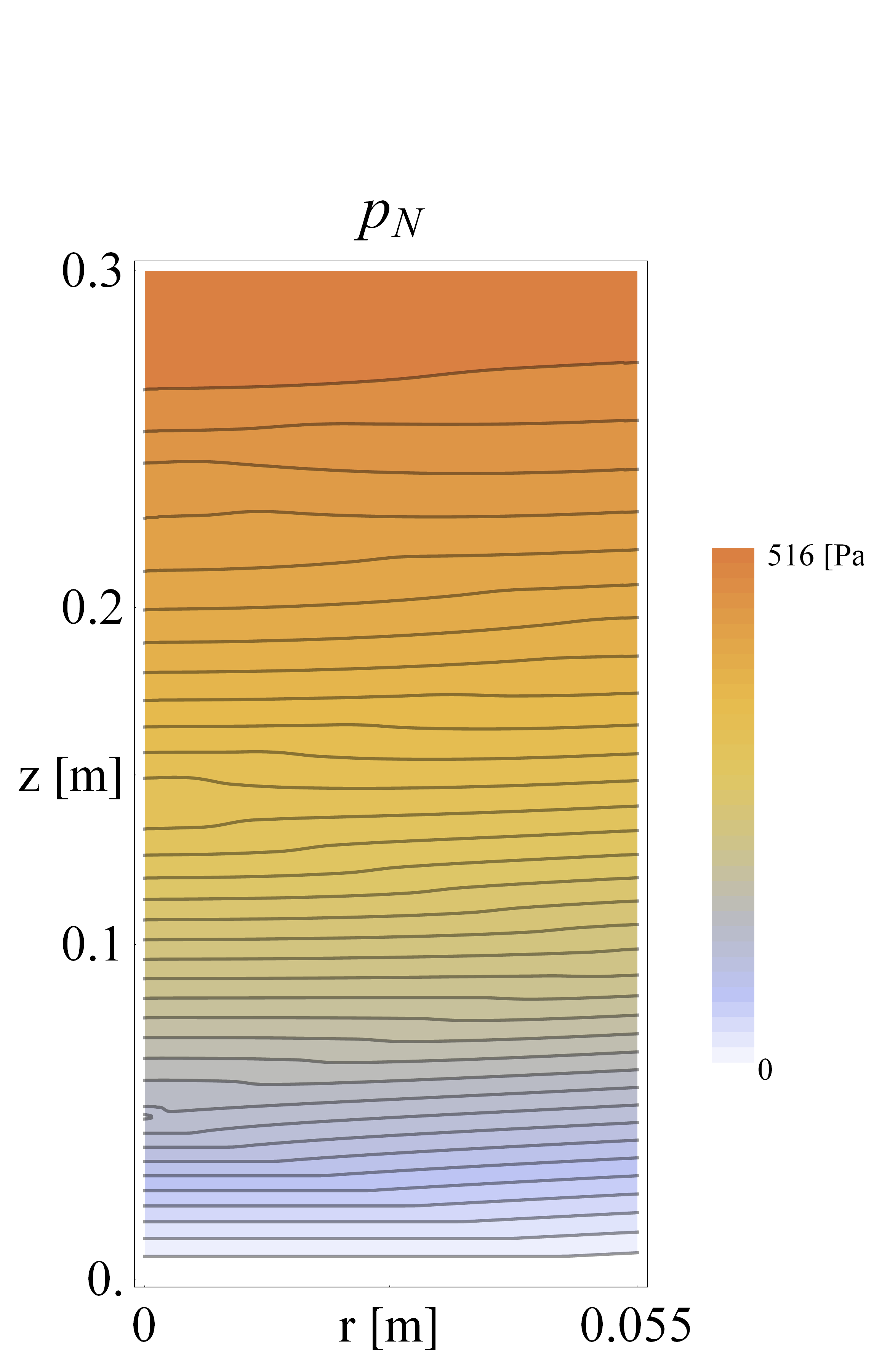}&
\includegraphics[width=0.3\columnwidth]{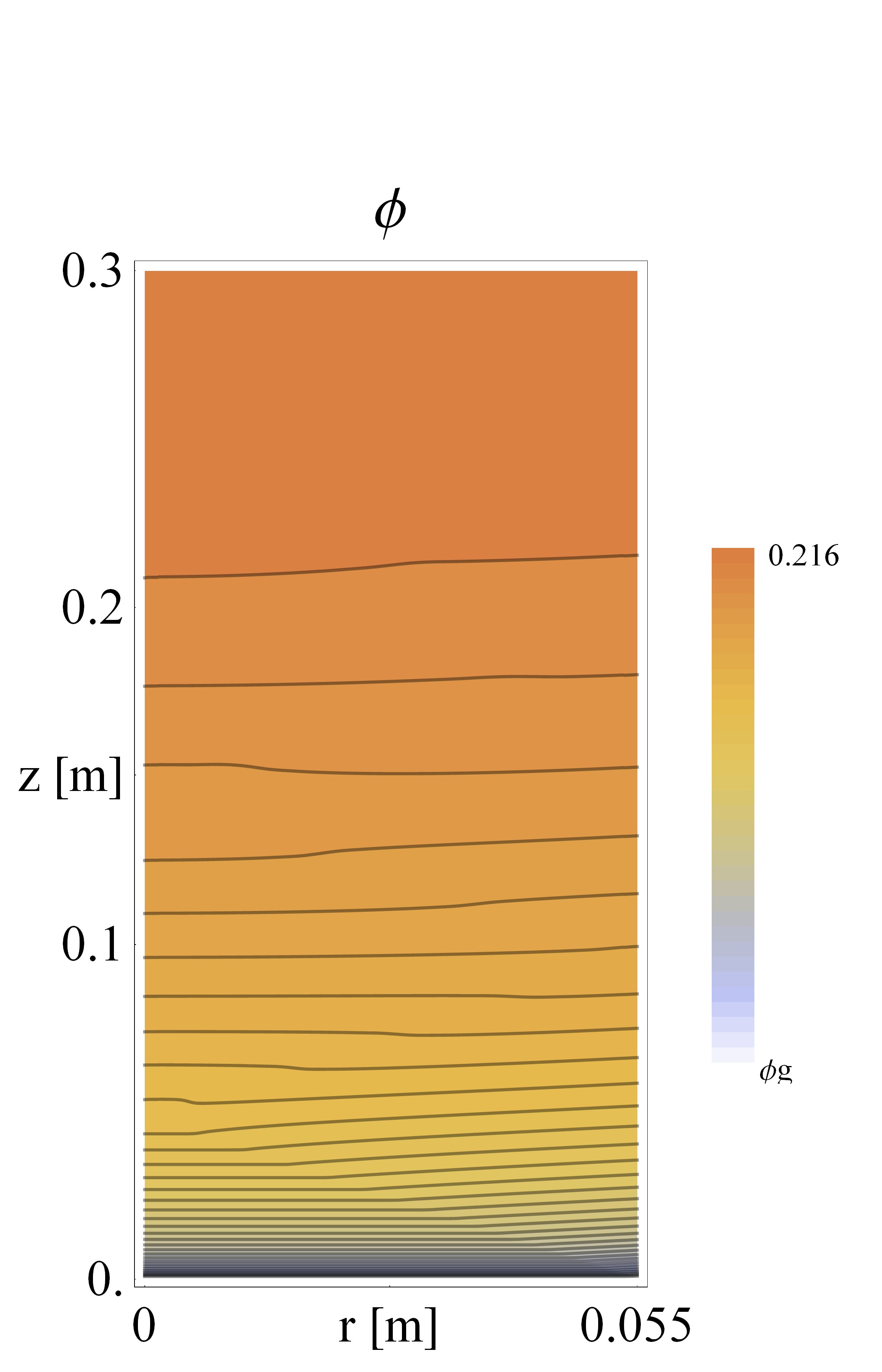}\\
(a) & (b) & (c)
\end{tabular}
\caption{Equilibrium (a) network shear stress $\tau_N$, (b) network pressure $p_N$, and (c) solids volume fraction $\phi$ distributions for suspension (a) in Table~\ref{tab:suspensions} as predicted by viscoplastic formulation in a 110 mm diameter container.}\label{fig:shearpress_large}
\end{figure}

Solutions to the coupled hyperbolic system (\ref{eqn:omega1}), (\ref{eqn:omega2}) are organized by characteristics which propagate from the top of the bed ($z=0$) at 45 degrees to the container wall and normal to each other as per Fig~\ref{fig:characteristics}(a), and manifest as $C^1$ shocks (i.e. non Lifshitz-continuous) contours in the shear stress distribution as shown in Fig~\ref{fig:characteristics}(b). Whilst the mathematical details differ, such 45 degree characteristics are typical of viscoplastic flows, such as slip-line fields in plasticity theory~\citep{Balmforth:2014,Hill:50}. The reflection conditions (\ref{eqn:r_zero_bc}), (\ref{eqn:r_R_bc}) correspond to the boundary conditions $\tau_N=0$ and $\tau_N=\tau_y(\phi)$ respectively, and $\omega_1$, $\omega_2$ increase from zero at the top of the bed (\ref{eqn:z_zero_bc}) due to the gravitational source in (\ref{eqn:omega1}) (\ref{eqn:omega2}). Predictions of the network shear stress, network pressure and solids volume fraction distributions for the colloidal suspension (a) in Table~\ref{tab:suspensions} in 22 mm and 110 mm diameter settling columns are shown in Fig.s~\ref{fig:shearpress_small} and \ref{fig:shearpress_large} respectively. The signature of the $\omega_1$, $\omega_2$ characteristics are clearly shown in the network shear stress distribution, and the number of times the characteristics are reflected from the $r=0$ and $r=R$ boundaries plays a significant role. The curved contours are due to the radial source terms $(\omega_1-\omega_2)/2r$ in (\ref{eqn:omega1}), (\ref{eqn:omega2}), whereas straight contours arise in Cartesian geometries.

For both columns represented in Fig.s~\ref{fig:shearpress_small} and \ref{fig:shearpress_large}, the network pressure and shear stress increase asymptotically with bed depth $z$ toward the equilibrium condition where all of the gravitational stress is borne by the shear stress. As such, the critical bed depth $z_c$ is never reached in practice, however it is possible to identify a finite representative depth $z_c'$ at which deviations from the equilibrium state are negligible. For the larger diameter column (Fig.~\ref{fig:shearpress_large}), the wider spacing between boundary reflections means that this equilibrium state is approached more slowly than in narrower columns. In general, these solutions of the equilibrium stress state appear to be physically plausible apart from the $C^1$ shocks in the network shear stress distribution, which arise from idealizations of the viscoplastic constitutive model and the closure approximation of zero normal stress differences. To resolve the accuracy of this viscoplastic approximation, the small strain hyperelastic formulation can be used to determine the true equilibrium state.

\section{Small Strain Solution of Hyperelastic Formulation}
\label{sec:small}

Central to solution of the equilibrium state $\zeta_\infty$ under the hyperelastic formulation are the assumptions stated above that (i) volumetric and isochoric strains for the suspension increase monotonically with time in an asymptotic fashion toward the equilibrium state, (ii) the equilibrium state is described by a critical state where the compressive yield stress balances the network pressure everywhere, and the network shear stress is equal to the shear yield strength at the container wall. These assumptions mean that the compressive irreversibility constraint need not be explicitly invoked, and similarly the shear yield criterion is never exceeded. As such, a unique equilibrium state is reached for all reasonable initial conditions where $\phi_0<\phi_g$.

This behaviour provides a significant simplification of the hyperelastic formulation, as it is no longer necessary to determine the path integral (\ref{eqn:hyperelastic}) from the initial condition $\zeta_0$, as any reasonable state which satisfies the conditions above shall ultimately converge to the unique equilibrium state $\zeta_\infty$. Hence, whilst the viscoplastic solution $\zeta_v$ may not represent a point on a true solution path from $\zeta_0$, under the hyperelastic formulation this state will still converge to the true equilibrium solution $\zeta_\infty$. The relationship between these states may be represented as
\begin{equation*}
\hat{\zeta_0}\xrightarrow{\mathbf{d}_v}\hat{\zeta_v}\Rightarrow\underset{\boldsymbol\kappa_v}{\zeta_v}\xrightarrow{\mathbf{d}_h}\zeta_\infty,
\end{equation*}
where the hat refers to the viscoplastic formulation, $\Rightarrow$ denotes conversion from the viscoplastic to the hyperelastic frame, and $\boldsymbol\kappa_v$ represents a reference state given by the hyperelastic frame. The total deformation $\mathbf{d}$ from the initial condition $\zeta_0$ in the frame $\boldsymbol\kappa_v$ is the sum of deformations under the viscoplastic and hyperelastic formulations
\begin{equation}
\mathbf{d}=\mathbf{d}_v+\mathbf{d}_h,\label{eqn:deformation}
\end{equation}
and total Hencky strain is given by the sum of the strains
\begin{equation}
\mathbf{H}=\mathbf{H}_v+\mathbf{H}_h.\label{eqn:Hdeformation}
\end{equation}


Due to the condition $\tau_N\leqslant \tau_y(\phi)$ and the brittleness of colloidal suspensions under shear, all deviatoric strains associated with relaxation from the viscoplastic frame to the hyperelastic frame are small $\gamma\leqslant\gamma_c$, and so in general the deviatoric Hencky strain is well-approximated by the infinitesimal strain tensor
\begin{equation}
\text{dev}(\mathbf{H})\approxeq\text{dev}(\boldsymbol\epsilon)=\text{dev}\left(\frac{1}{2}(\nabla_v\mathbf{d}+(\nabla_v\mathbf{d})^T)\right),
\end{equation}
with $\gamma\approxeq\sqrt{\text{dev}(\boldsymbol\epsilon):\text{dev}(\boldsymbol\epsilon)}$. Conversely, whilst the deviatoric strains are small, the particulate network can undergo large-scale consolidation, and hence support large isotropic strains, the volumetric component of the finite strain measure (\ref{eqn:K1}) in the material frame $\boldsymbol\kappa_v$ must be preserved as
\begin{equation}
\phi=\phi_0\exp(-\nabla\cdot\mathbf{d}),\label{eqn:phi_d}
\end{equation}
and so the total Hencky strain for strongly flocculated colloidal gels may be well approximated as
\begin{equation}
\mathbf{H}=\frac{1}{3}\ln\left(\frac{\phi_0}{\phi}\right)\mathbf{I}+\text{dev}(\boldsymbol\epsilon).
\end{equation}

Under these strain measures, the suspension network pressure (\ref{eqn:hyperpressure}) and deviatoric stress (\ref{eqn:hypershear}) are now
\begin{align}
&p_N=\int_{\phi_0}^{\phi(t)}K(\phi)d\ln\phi= P_y(\phi_v)+\int_{\phi_v}^{\phi(t)}K(\phi)d\ln\phi,\label{eqn:hyperpressuresmall}\\
&\boldsymbol\sigma_N=\int_{-\infty}^t \frac{\partial G(\phi,\gamma,t-s)}{\partial s}\text{dev}(\boldsymbol\epsilon(s)) ds\label{eqn:hypershearsmall}.
\end{align}
As the timescale of material deformations of the particulate network is slow (due to the magnitude of the interphase drag coefficient $R(\phi)$), the shear modulus $G(\phi,\gamma,t)$ is well-approximated by the infinite-time modulus $G_\infty(\phi,\gamma)=\lim_{t\rightarrow\infty}G(\phi,\gamma,t)$. Furthermore, as the critical shear strain $\gamma_c$ is small, it is unnecessary to resolve the nonlinear shear strain prior to yield, and so the infinite-time modulus may be linearized from (\ref{eqn:tau_G}) as
\begin{equation}
G_\infty(\phi)=\frac{1}{\gamma_c}\tau_y(\phi),
\end{equation}
and integration by parts of (\ref{eqn:hypershearsmall}) yields
\begin{equation}
\begin{split}
\boldsymbol\sigma_N=\int_{-\infty}^t G_\infty(\phi)\frac{\partial}{\partial s}\text{dev}(\boldsymbol\epsilon(s))ds\label{eqn:hypershearexplicit}.
\end{split}
\end{equation}


In converting the viscoplastic solution $\hat{\zeta}_v$ to the hyperelastic formulation $\zeta_v$, this solution is now over-determined in the hyperelastic frame due to the closure $N_1=N_2=0$ developed in Section~\ref{sec:closure}. Relaxation of this constraint perturbs the force balance (\ref{eqn:eqmbalance}) away from equilibrium, and so the relaxed viscoplastic solution forms an initial condition for the evolution equation (\ref{eqn:evoln}) which can then evolve to the true equilibrium state. This process can be formally represented as
\begin{equation}
\nabla\cdot\hat{\Sigma}^{N,v}+\Delta\rho\mathbf{g}\phi_v=0\Rightarrow\nabla\cdot\Sigma^{N,v}+\Delta\rho\mathbf{g}\phi_v=\mathbf{S},\label{eqn:stress_imbalance}
\end{equation}
where $\mathbf{S}$ is the force imbalance due to conversion of the network stress $\hat{\Sigma}^{N,v}$ in the viscoplastic frame to the network stress $\Sigma^{N,v}$ in the hyperelastic frame. This conversion requires calculation of the deformation vector $\mathbf{d}_v$ from the solids volume fraction distribution $\phi_v$ as
\begin{equation}
\nabla\cdot\mathbf{d}_v=\log\left(\frac{\phi_0}{\phi_v}\right),\label{eqn:phi_dv}
\end{equation}
subject to the boundary condition $\mathbf{d}_v|_{r=0,z=0}=0$. Although it is not possible to determine the temporal evolution of $\mathbf{d}$ up to the viscoplastic solution, the arguments above justify that convergence to the unique equilibrium state are independent of the solution path. Hence, for simplicity we assume $\mathbf{d}$ evolves linearly under the viscoplastic solution as
\begin{equation}
\mathbf{d}(t)=h(t)\mathbf{d}_v,\quad\text{for}\,\,\,\, t\leqslant t_v,
\end{equation}
where $h(t)$ is the ramp function
\begin{equation}
h(t)=
\begin{cases}
&0\quad t\leqslant 0,\\
&t/t_v\quad 0 < t \leqslant t_v,\\
&1 \quad t_v<t,
\end{cases}
\end{equation}
and $t_v$ is a nominal time at which the viscoplastic solution occurs. As such, the deviatoric network stress at $t_v$ is
\begin{equation}
\boldsymbol\sigma_N^v=\int_0^{t_v}G_\infty(\phi(s))h'(s)\text{dev}(\boldsymbol\epsilon_v)ds=H(\phi_v)\frac{\text{dev}(\boldsymbol\epsilon_v)}{\nabla\cdot\mathbf{d}_v},
\end{equation}
where $H(\phi_v)=\int_{\phi_0}^{\phi_v}\frac{1}{\varphi}G_\infty(\varphi)d\varphi$. Hence the stress imbalance arising from relaxation of the closure constraint manifests as
\begin{equation}
\mathbf{S}=\nabla\cdot\left(\Sigma^{N,v}-\hat{\Sigma}^{N,v}\right)=\nabla\cdot\left(H(\phi_v)\frac{\text{dev}(\boldsymbol\epsilon_v)}{\nabla\cdot\mathbf{d}_v}-\hat{\tau}_N^v\left(
\begin{array}{cc}
0 & 1\\
1 & 0
\end{array}
\right)\right),
\end{equation}
where $\hat{\tau}_N^v$ is the network shear stress calculated for the viscoplastic solution.


As the total network stress is $\Sigma^N=\Sigma^{N,v}+\Sigma^{N,h}$, where $\Sigma^{N,h}$ is the hyperelastic component of the total stress, then from (\ref{eqn:stress_imbalance}) the evolution equation (\ref{eqn:evoln}) from the viscoplastic solution $\zeta_v$ is
\begin{equation}
\frac{\partial\phi}{\partial t}+\mathbf{q}\cdot\nabla\phi+\nabla\cdot\left[\frac{(1-\phi)^2}{R(\phi)}\left(\nabla\cdot\Sigma^{N,h}+\Delta\rho\mathbf{g}\delta\phi+\mathbf{S}\right)\right]=0,\label{eqn:hyperevol}
\end{equation}
where $\delta\phi=\phi-\phi_v$, $\phi|_{t=t_v}=\phi_v$, and $\Sigma^{N,h}|_{t=t_v}=\mathbf{0}$. The solids phase velocity is defined as $\mathbf{v}_s\equiv\frac{\partial}{\partial t}\mathbf{d}$, and the evolution equation (\ref{eqn:evoln}) is of the form~\citep{LesterEA:10}
\begin{equation}
\frac{\partial\phi}{\partial t}+\mathbf{q}\cdot\nabla\phi+\nabla\cdot((1-\phi)\mathbf{v}_r)=0,\label{eqn:solids}
\end{equation}
where $\mathbf{v}_r=\frac{\mathbf{v}_s-\mathbf{q}}{1-\phi}$ is the inter-phase velocity. Under the assumption $\mathbf{q}=0$, (\ref{eqn:hyperevol}), (\ref{eqn:solids}) may be re-cast as an evolution equation for the hyperelastic deformation $\mathbf{d}_h$
\begin{equation}
\frac{\partial}{\partial t}\mathbf{d}_h=\frac{(1-\phi)^2}{R(\phi)}\left(\nabla\cdot\Sigma^{N,h}+\Delta\rho\mathbf{g}\delta\phi+\mathbf{S}\right),\label{eqn:devol}
\end{equation}
subject to the initial condition $\mathbf{d}_h|_{t=t_v}=\mathbf{0}$ and the same boundary conditions (\ref{eqn:wallcondition})-(\ref{eqn:topcondition}) as for the viscoplastic problem. The network tensor $\Sigma^{N,h}$ is explicitly
\begin{equation}
\Sigma^{N,h}=\int_{\phi_v}^{\phi}K(\varphi)d\ln\varphi\mathbf{I}+\int_{t_v}^{t}G_\infty(\phi)\frac{\partial}{\partial s}\text{dev}(\boldsymbol\epsilon_h) ds,
\end{equation}
where the bulk and shear moduli are given from $P_y(\phi)$, $\tau_y(\phi)$, $\gamma_c$ in (\ref{eqn:Sphi}), (\ref{eqn:pyphi_func}) as
\begin{align}
&K(\phi)=\frac{1}{\phi}\frac{d P_y(\phi)}{d\phi},\\
&G_\infty(\phi)=\frac{1}{\gamma_c}\tau_y(\phi).
\end{align}

\begin{figure}
\centering
\begin{tabular}{c c c}
\includegraphics[width=0.3\columnwidth]{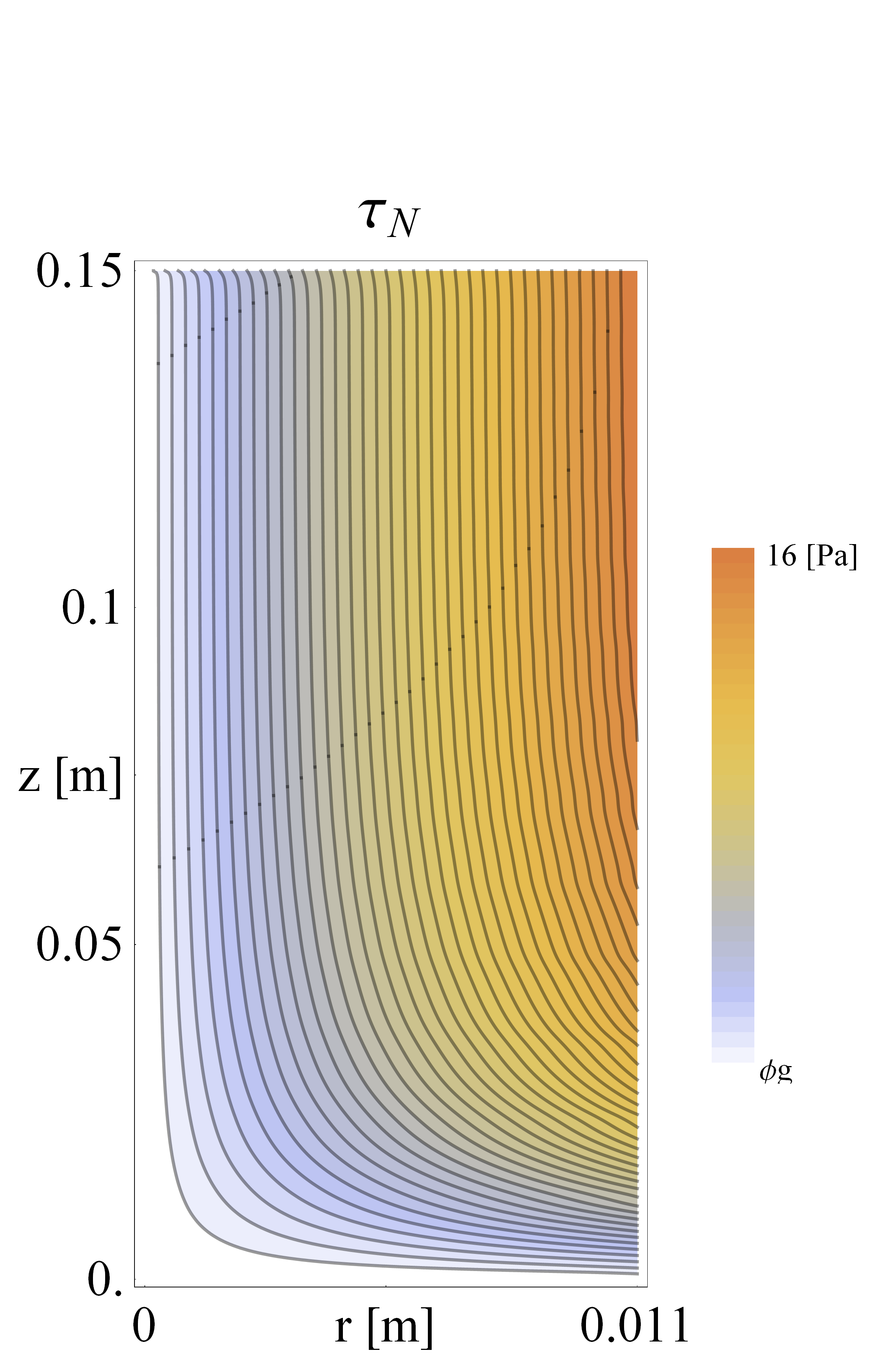}&
\includegraphics[width=0.3\columnwidth]{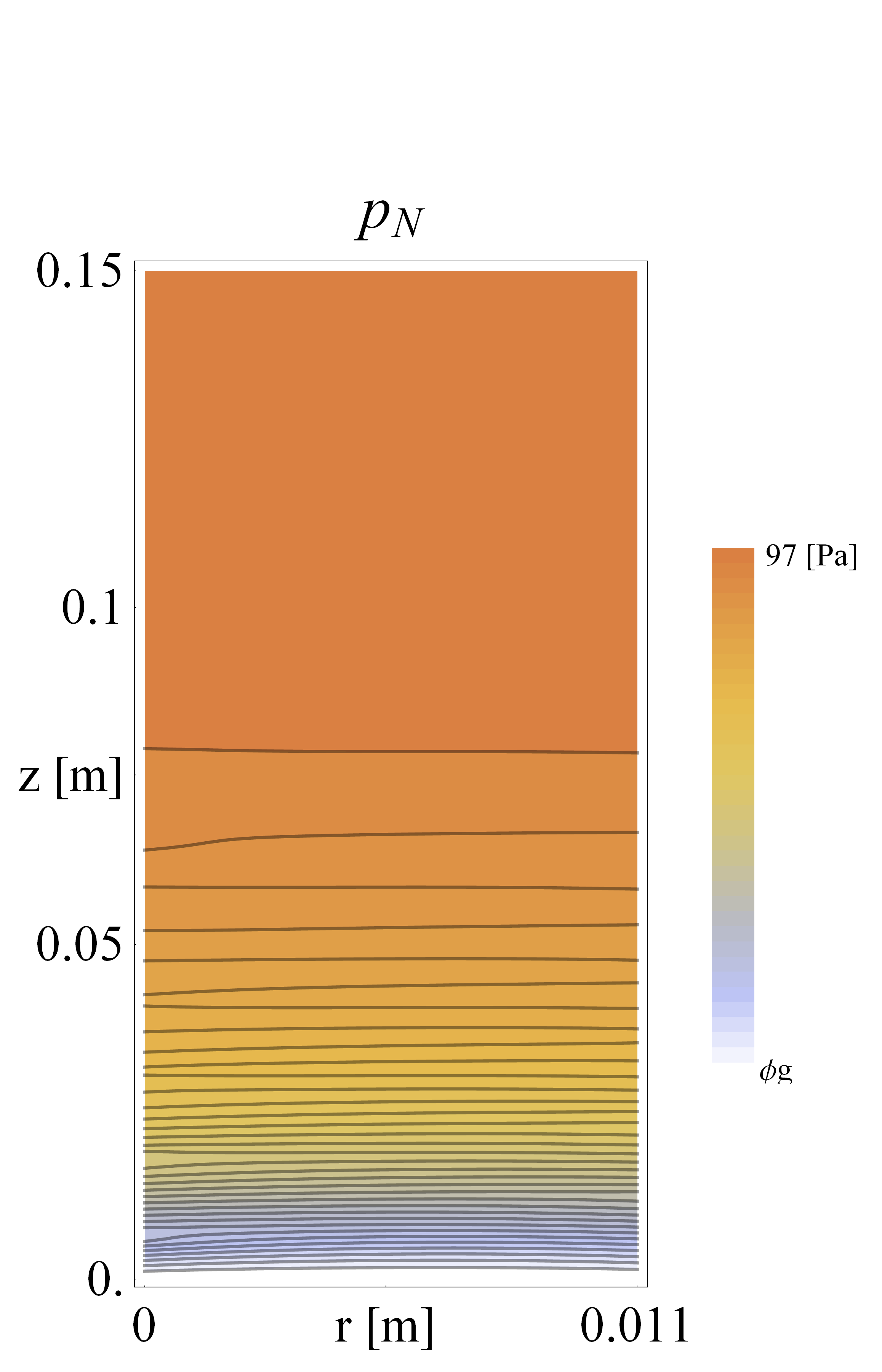}&
\includegraphics[width=0.3\columnwidth]{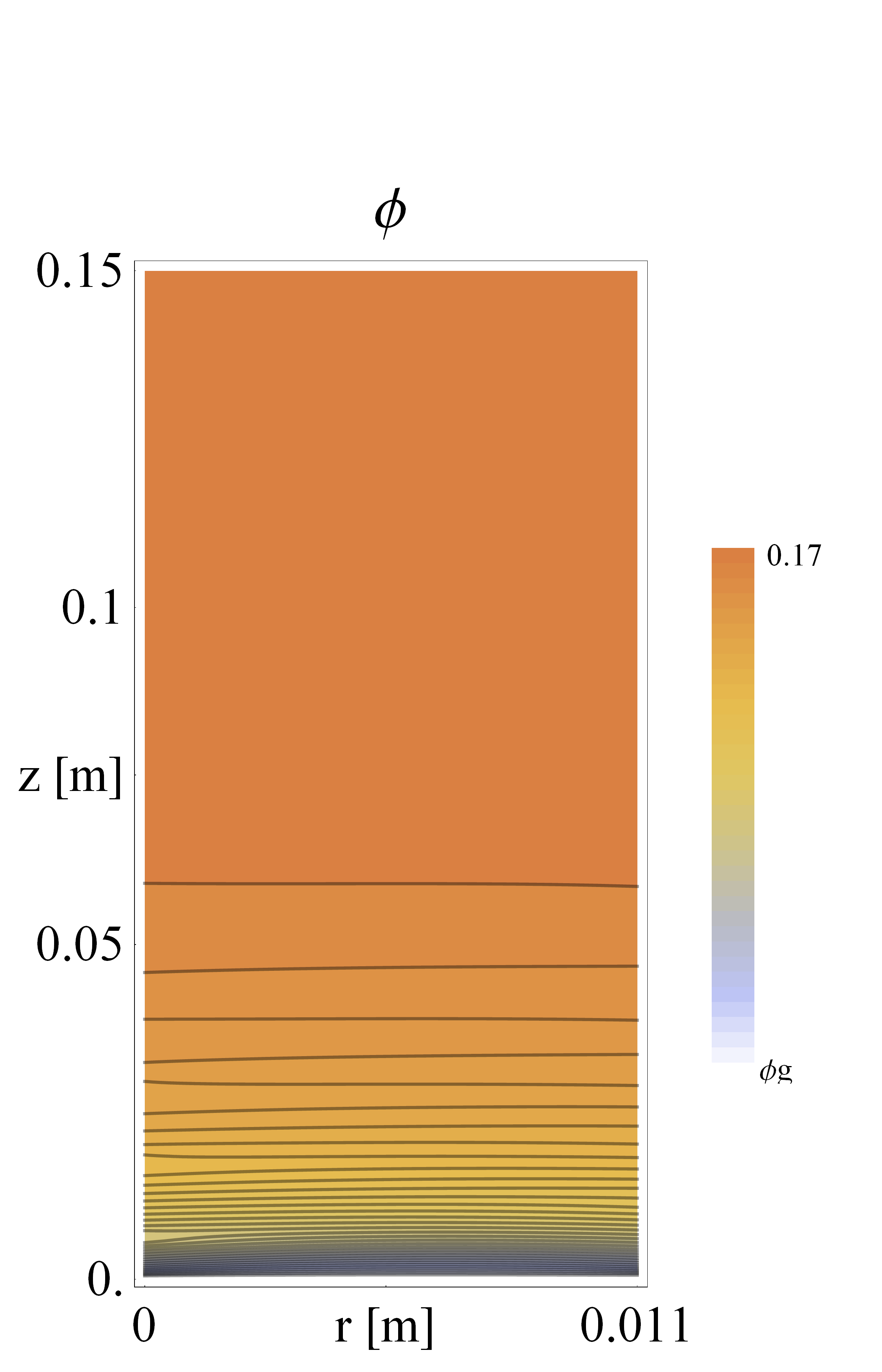}\\
(a) & (b) & (c)
\end{tabular}
\caption{Equilibrium (a) network shear stress $\tau_N$, (b) network pressure $p_N$, and (c) solids volume fraction $\phi$ distributions for suspension (a) in Table~\ref{tab:suspensions} as predicted by the hyperelastic formulation in a 22 mm diameter container.}\label{fig:shearpress_small_hyper}
\end{figure}

\begin{figure}
\centering
\begin{tabular}{c c}
\includegraphics[width=0.3\columnwidth]{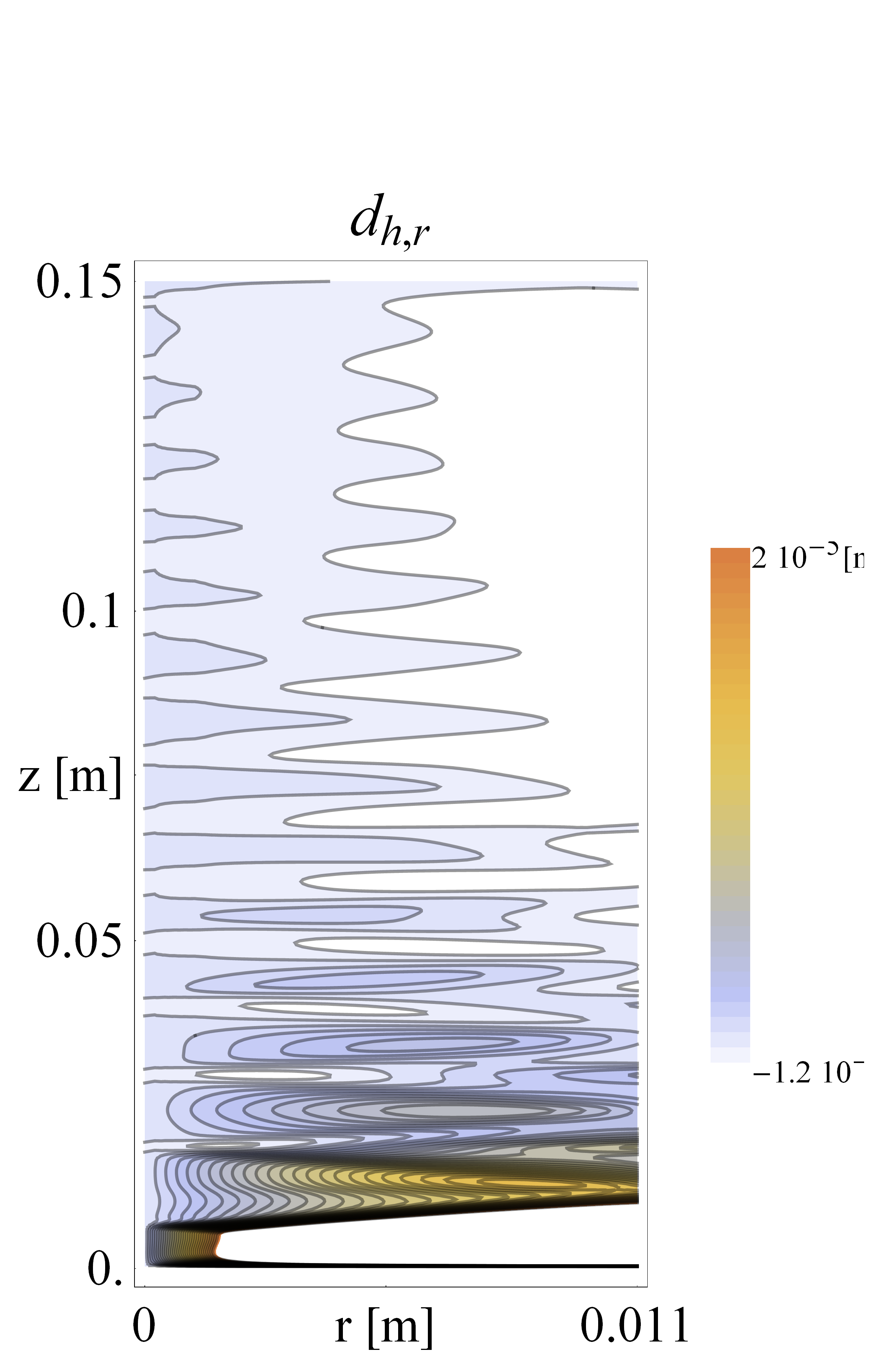}&
\includegraphics[width=0.3\columnwidth]{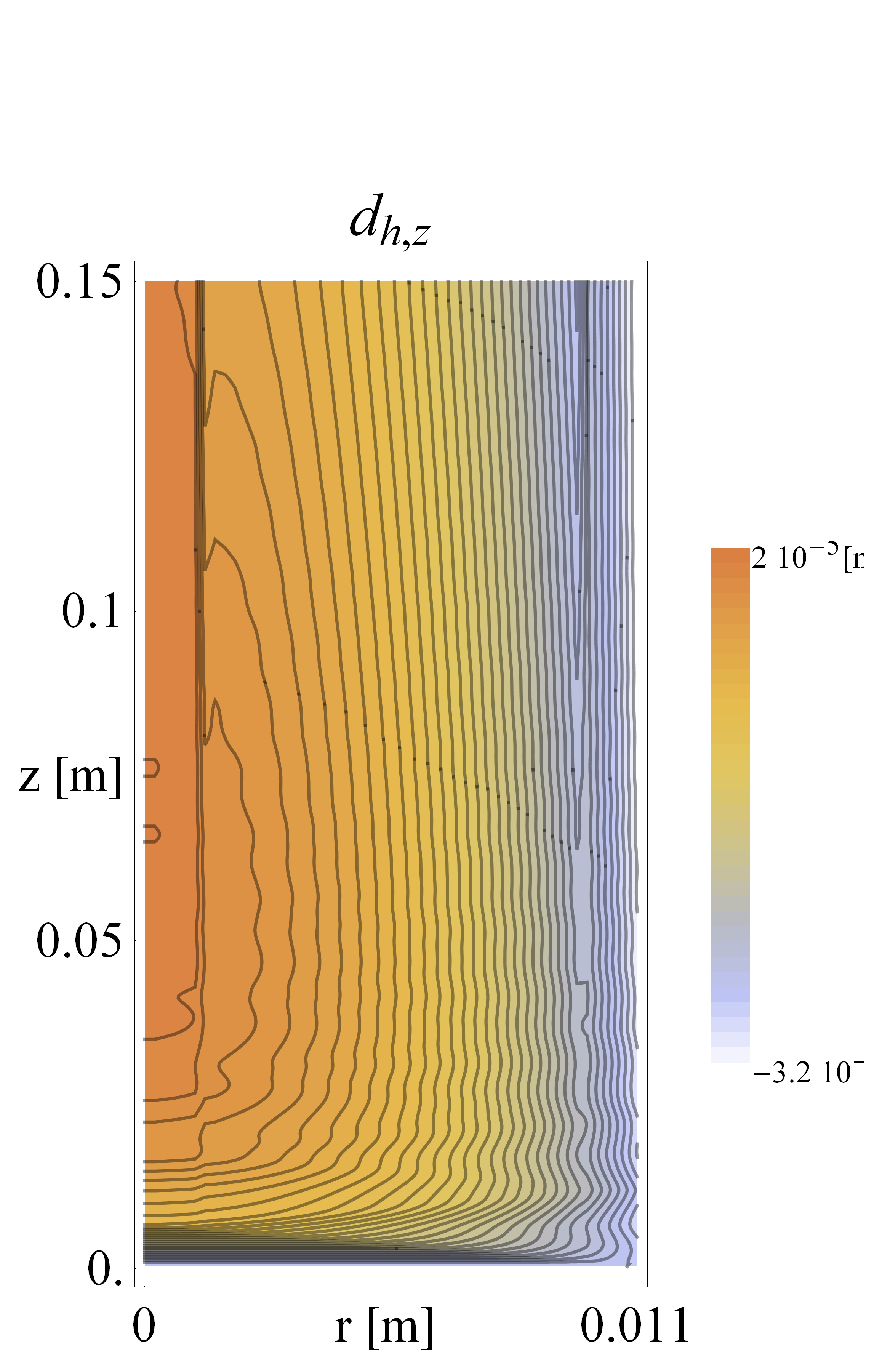}\\
(a) & (b)
\end{tabular}
\caption{distributions of (a) radial $d_{h,r}$ and (b) vertical $d_{h,z}$ components of the hyperelastic deformation $\mathbf{d}_h$ as predicted by (\ref{eqn:deformation}) in a 22 mm diameter container.}\label{fig:hyper_strain}
\end{figure}

The evolution equation (\ref{eqn:devol}) describes relaxation of the particulate network from the viscoplastic solution to the hyperelastic equilibrium condition as a nonlinear elliptic equation. A finite difference routine is used to numerically solve (\ref{eqn:devol}) on a series of increasingly fine spatial grids subject to successive over-relaxation to aid both robustness and convergence. Under this routine, (\ref{eqn:devol}) converges relatively quickly, and the equilibrium distributions of $\tau_N$, $p_N$ and $\phi_N$ for suspension (a) in the 22 mm column are shown in Fig.~\ref{fig:shearpress_small_hyper} (a)-(c), with the associated hyperelastic deformation $\mathbf{d}_h$ shown in Fig.~\ref{fig:hyper_strain}. The most striking impact of the hyperelastic equilibrium compared to Fig.~\ref{fig:shearpress_small} is the smoothing of the $C^1$ shocks from the network shear stress distribution, resulting in a smooth radial profile of $\tau_N$. This removal is related to the change in governing equations from hyperbolic to elliptic, and demonstrates clearly that the $C^1$ shocks are spurious artifacts of the viscoplastic formulation under the closure approximation $N_1=N_2=0$. Whilst the smoothing of $C^1$ shocks appears markedly in Fig.~\ref{fig:shearpress_small_hyper}(a), the $L_2$ norm of $\tau_N$ between the viscoplastic and hyperelastic solutions is of the order 2\%, and so the gross distribution of shear stress is preserved. Likewise the network pressure and solid volume fraction distributions are only slightly altered ($L_2$ norm 0.5\%, 0.3\% respectively). The magnitude of the hyperelastic strain is also small, $|\mathbf{d}_h|\leqslant 3.2\times10^{-6}$ m. From Fig.~\ref{fig:hyper_strain}, these strains are primarily located near the $C^1$ shocks, and so most of the deformation relaxation of the hyperelastic body occurs as isochoric strains due to the stress imbalance generated by these shocks. As such, whilst the viscoplastic formulation under the closure assumption $N_1=N_2=0$ does introduce spurious artifacts in the form of the reflected $C^1$ shocks in the shear stress field, this model still yields accurate estimates of the solids volume fraction $\phi$ and network pressure $p_N$ distributions, and captures the gross features of the shear stress $\tau_N$ distribution. Despite this deficiency, this model serves as a useful tool for the modeling and characterization of sedimentation in the presence of significant wall adhesion effects. Most importantly, the hyperelastic solution above directly quantifies the nature and extent of the errors associated with the viscoplastic solution.

\section{Application of Viscoplastic Solution to Experimental Data}
\label{sec:application}

Solution of the small-strain hyperelastic formulation above shows that the viscoplastic formulation under the closure assumption ($N_1=N_2=0$) for wall adhesion in batch sedimentation yields an accurate approximation of the solids volume fraction $\phi$ and network pressure $p_N$ distributions, and captures the gross features of the network shear stress $\tau_N$. This represents a significant simplification, as the hyperelastic formulation (\ref{eqn:hyperelastic}) (\ref{eqn:evoln}) is significantly more computationally intensive, even if the small-strain hyperelastic formulation (\ref{eqn:devol}) is invoked. To test the viscoplastic model and develop methods to extract the relevant rheological functions, we apply the viscoplastic model to equilibrium solids volume fraction profile data (measured using a gamma-ray attenuation device~\citep{LabbettEA:06}) across batch settling columns of various widths to estimate both the compressive and shear yield strength functions. These estimates are then compared to \emph{in-situ} measurements of the shear yield stress using a vane rheometer, which serves as an indirect test of the viscoplastic model.

\begin{figure}
\begin{centering}
\begin{tabular}{c c c}
\includegraphics[width=0.32\columnwidth]{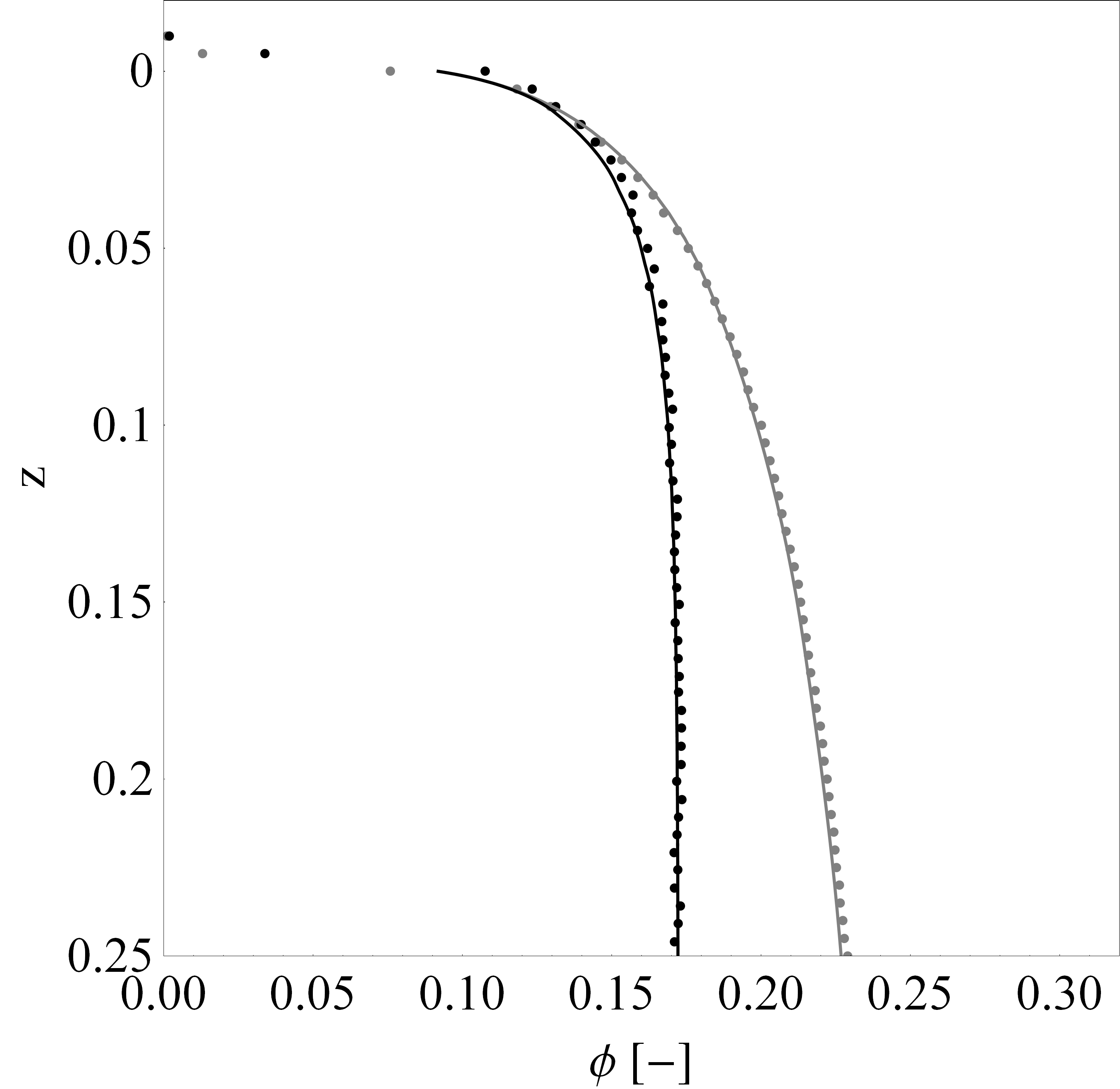}&
\includegraphics[width=0.32\columnwidth]{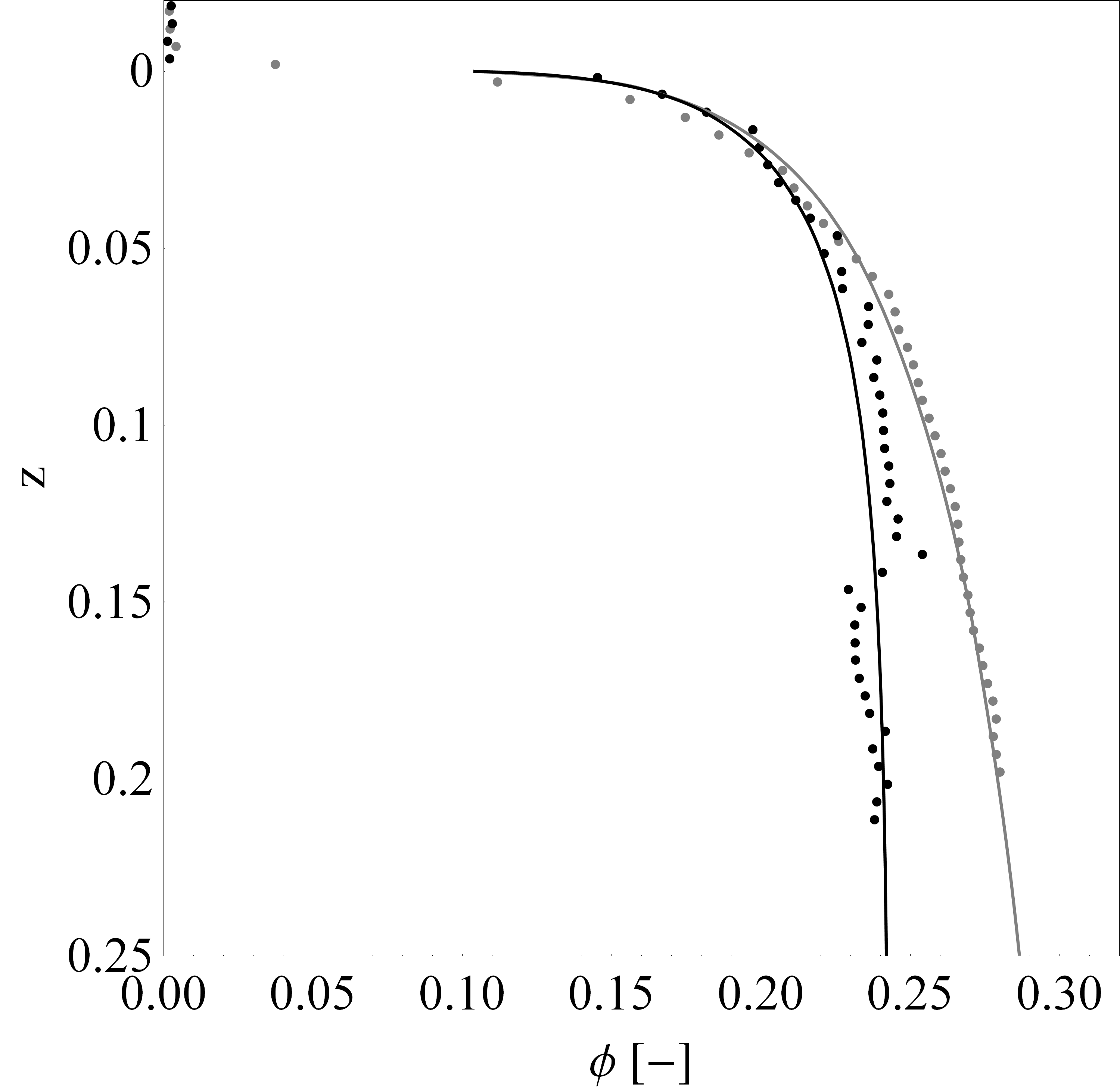}&
\includegraphics[width=0.32\columnwidth]{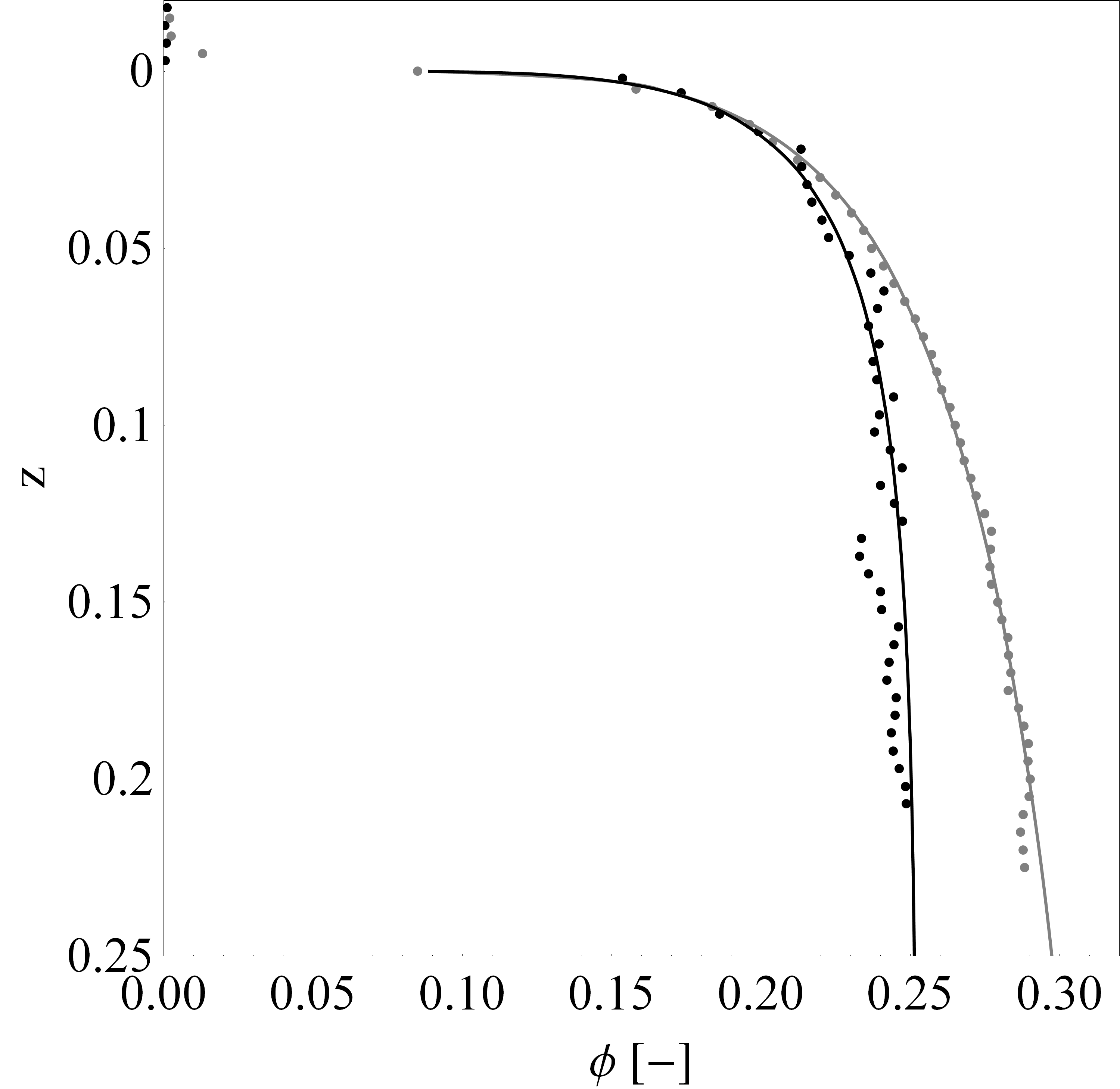}\\
(a) & (b) & (c)
\end{tabular}
\end{centering}
\caption{Measured data (points) and model predictions (lines) of the equilibrium solids volume fraction profile $\phi_\infty$ for $R_s$=0.011[m] (black) and $R_l$=0.055[m] (gray) column widths for calcium carbonate suspensions under flocculant types and dosages (a)-(c) summarized in Table~\ref{tab:suspensions}.}\label{fig:profiles_fitted} 
\end{figure}

In this study we consider three suspensions which consist of mean size 4 $\mu$m calcium carbonate primary particles (Omyacarb 2, Omya Australia Pty Ltd.) of density difference $\Delta\rho=1720 $kg m$^{-3}$ with respect to an aqueous solution. The primary particles are flocculated in a 22 mm ID continuous flow pipe reactor~\citep{OwenEA:08} using two different commercial high molecular weight polymer flocculants (Magnafloc 336 and Rheomax DR 1050, BASF). Both flocculants were made up at 0.1 w/w\% and diluted immediately prior to use to 0.01 w/w\% prior to application at a solids concentration of 90 g/L for a residence time of 9.9 s in the continuous flow pipe reactor operating at a flowrate of 20 L/min. Further details of the flocculation conditions are described by \citet{OwenEA:08}, and the different dosages and remaining flocculation conditions are summarized in Table~\ref{tab:suspensions}. The settling behaviour of these three suspensions was measured in two different diameter (22  mm and 110 mm) cylindrical settling columns of initial height $h_0\approx 2000$ mm and initial solids concentration $\phi_0\approx 0.033$. These suspensions exhibit markedly different transient and equilibrium sedimentation behaviour, as reflected in equilibrium solids volume fraction profile data under batch settling (shown in Fig.~\ref{fig:profiles_fitted}), as measured by gamma ray attenuation in both 22 mm and 110 mm diameter cylindrical settling columns. Evidence of wall adhesion is clearly shown in the 22 mm columns, where the solids volume fraction profile evolves to a constant value (within scatter) with increasing bed depth, suggesting that beyond a critical depth the suspension weight is entirely supported by the container walls.

\begin{table}[b]
\begin{centering}
\begin{tabular}{c c c c c c c}
\hline
Suspension & Flocculant & Dosage [g/t] & $\phi_g$ [-] & $k$ [Pa] & $n$ [-] & $S_\infty$ [-] \\
\hline
(a) & Magnafloc 336 & 46 & 0.0918 & 3.21 & 5.48 & 0.157 \\
(b) & Rheomax DR 1050 & 30 & 0.1042 &  0.63 & 7.03 & 0.112 \\
(c) & Rheomax DR 1050 & 46 & 0.0890 & 0.16 & 7.01 & 0.113 \\
\hline
\end{tabular}
\caption{Suspension flocculant type, dosage and fitted rheological parameters.}\label{tab:suspensions}
\end{centering}
\end{table}

\begin{centering}
\begin{figure}
\centering
\includegraphics[width=0.65\columnwidth]{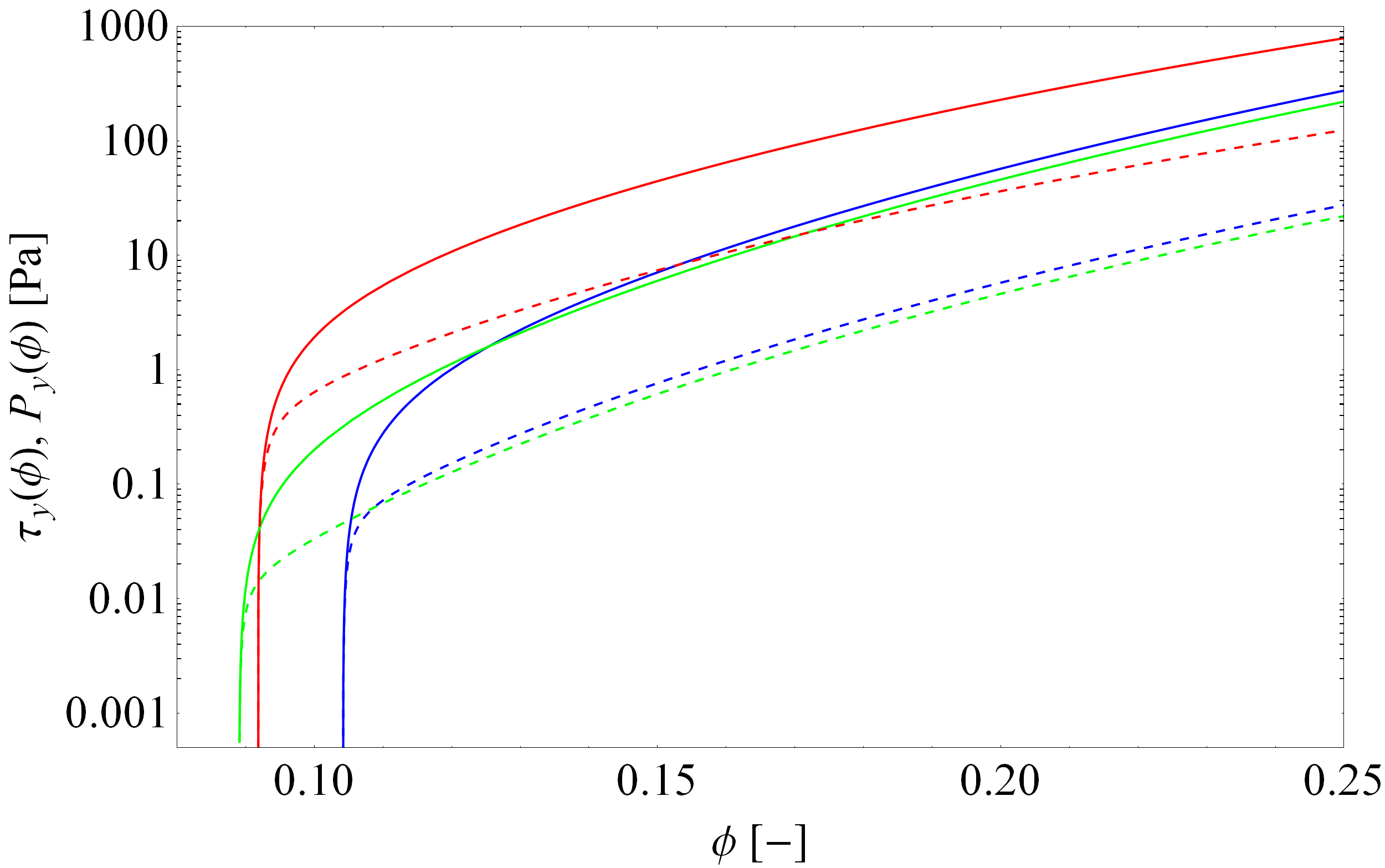}
\caption{Fitted compressive yield strength $P_y(\phi)$ (solid) and shear yield strength $\tau_y(\phi)$ (dashed) curves from viscoplastic model for suspensions (a) (red), (b) (blue), (c) (green) summarized in Table~\ref{tab:suspensions}.}\label{fig:pytau_fitted}
\end{figure}
\end{centering}

Estimation of the compressive and shear yield stress functions is performed by fitting of the relevant rheological parameters via minimization of the $L_2$ error between the solids volume fraction profiles as predicted by the viscoplastic model and the measured data over the 22 mm and 110 mm columns. The functional form (\ref{eqn:pyphi_func}) is used for the compressive yield strength function $P_y(\phi)$, and the critical strain relationship proposed by \citet{Buscall:09} is used for the shear yield strength function $\tau_y(\phi)$, where $\tau_y(\phi)=S(\phi)P_y(\phi)$, and $S(\phi)$ is given by (\ref{eqn:Sphi}). Hence fitting of the equilibrium solids volume fraction profiles comprises of four rheological constants: the suspension gel point $\phi_g$, the consistency $k$ in (\ref{eqn:pyphi_func}), the index $n$ in (\ref{eqn:pyphi_func}), and the asymptotic shear/compressive yield strength ratio $S_\infty$ in (\ref{eqn:Sphi}). Note that the suspension gel point can be accurately estimated directly from the equilibrium solids volume fraction data $\emph{a priori}$, and so the numerical fitting method only involves 3 variable parameters. Minimization of the $L_2$ error between model predictions and experimental data is performed via a simplex optimisation routine, where the numerical resolution of the finite difference routine used to solve the viscoplastic model (\ref{eqn:omega1}) (\ref{eqn:omega2}) is increased as the rheological parameters converge. Although the viscoplastic model predicts variation in the radial network pressure distribution (e.g. Fig.~\ref{fig:shearpress_small_hyper} (c)), the resultant variation in solids volume fraction profile is weak, and so such variations are neglected in comparison between experimental data and model predictions. The vertical solids volume fraction profiles from the fitted viscoplastic model are shown in Fig.~\ref{fig:profiles_fitted} along with the experimental gamma ray attenuation, and the fitted compressive and shear yield strength curves are shown in Fig.~\ref{fig:pytau_fitted}. The rheological parameters associated with these fits are also summarized in Table~\ref{tab:suspensions}, of note is the large asymptotic yield strength ratio $S_\infty\approx 0.1-0.15$ required to fit the measured solids volume fraction profiles. We found no means of fitting the experimental data in Fig.~\ref{fig:profiles_fitted} without incorporating such values of $S_\infty$, regardless of the functional form of $\tau_y(\phi)$, $P_y(\phi)$, hence we conclude that these values are an accurate representation of the asymptotic yield strength ratio, independent of fitting methodology.

The range $S(\phi)\approx$ 0.1-1 found here for the materials summarised in Table~\ref{tab:suspensions} is overall significantly larger than the range $S_\infty\approx$ 0.001-0.2 reported previously~\citep{BuscallEA:87,BuscallEA:88,deKretserEA:02,ZhouEA:01,Channell/Zukoski:97} for a range of strongly flocculated suspensions (and summarised in \citet{Buscall:09}), leading to much stronger wall effects. Overall, $S(\phi)$ is expected to decay from unity at the gel-point towards the asymptotic value $S_\infty$ dependent \emph{inter alia} upon the elasticity of the interparticle bonds and the particle-size~\citep{Buscall:09}. Much of the earlier data refers to electrolyte-coagulated systems away from the gel-point and, with the benefit of hindsight, it is perhaps not too surprising to find that high-polymer flocculated systems are different. Furthermore, more than a dozen types or mechanisms of flocculation are known and this alone means that wall effects are likely to be more important for some systems than others.


\begin{centering}
\begin{figure}
\includegraphics[width=0.45\columnwidth]{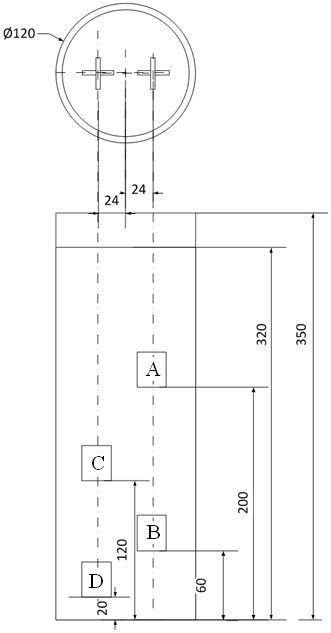}
\centering
\caption{Schematic of shear yield stress measurement protocol in the 110 mm ID column using an \emph{in-situ} vane rheometer. The top schematic shows the vane placement with respect to the column cross-section, and the bottom schematic shows a side profile of the vane locations at different heights of the column. All dimensions are in mm, and the vane measurement order is given by $A$-$D$. Measured data is shown in Table~\ref{tab:rheometry}}\label{fig:vane_schematic}
\end{figure}
\end{centering}

\begin{table}[b]
\begin{centering}
\begin{tabular}{c c c c c c c}
\hline
Suspension & Flocculant & Dosage [g/t] & A [Pa] & B [Pa] & C [Pa] & D [Pa]\\
\hline
(a) & Magnafloc 336 & 46 & 14.58 & 80.68 & 61.24 & 95.01 \\
(b) & Rheomax DR 1050 & 30 & 4.37 & 41.55 & 26.97 & 56.38 \\
(c) & Rheomax DR 1050 & 46 & 5.35 & 49.33 & 27.46 & 57.83 \\
\hline
\end{tabular}
\caption{Table of \emph{in-situ} vane shear yield strength measurements for vane protocol A-D shown in Fig.~\ref{fig:vane_schematic}}\label{tab:rheometry}
\end{centering}
\end{table}

\begin{figure}
\begin{centering}
\begin{tabular}{c c c}
\includegraphics[width=0.32\columnwidth]{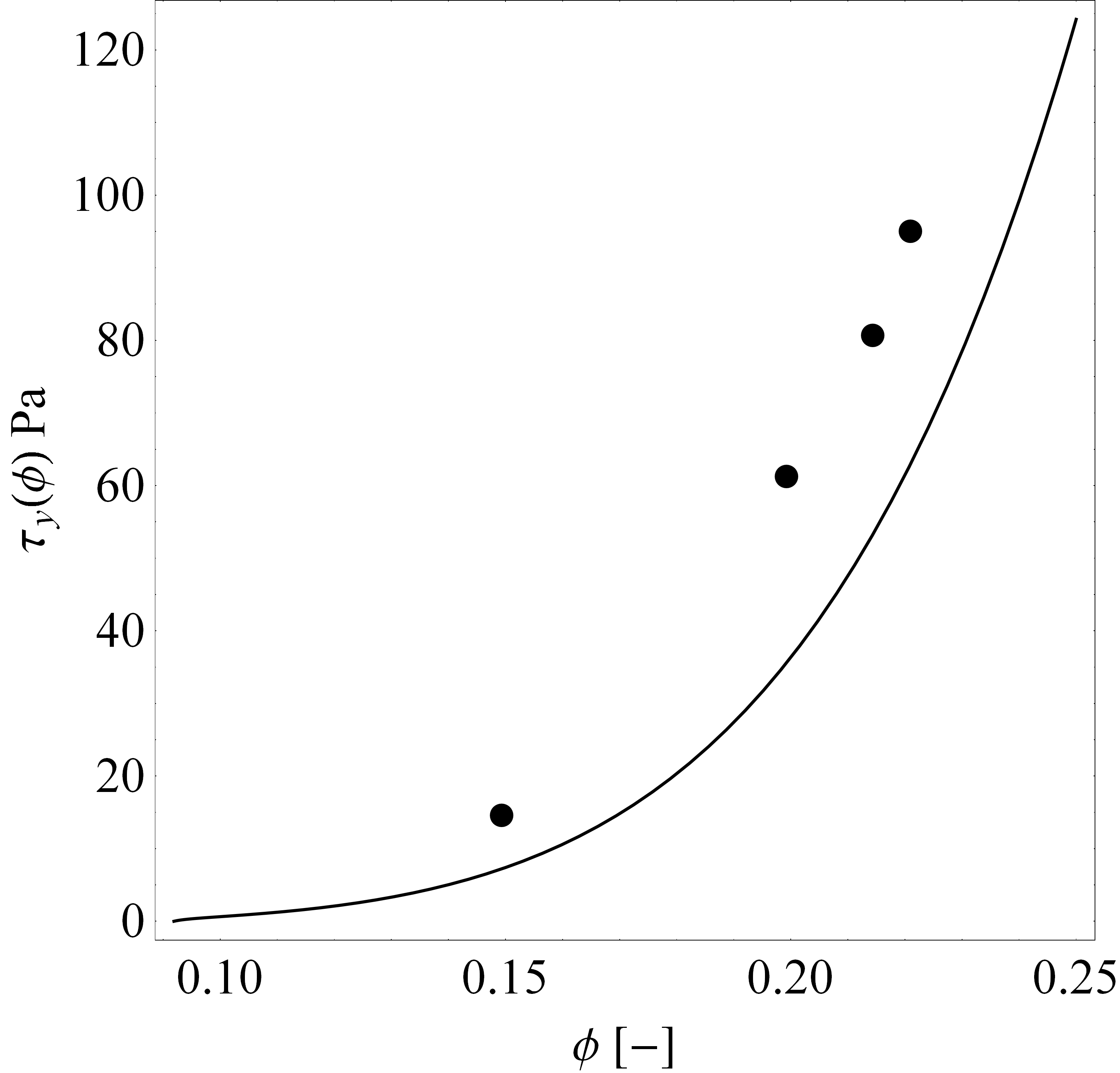}&
\includegraphics[width=0.32\columnwidth]{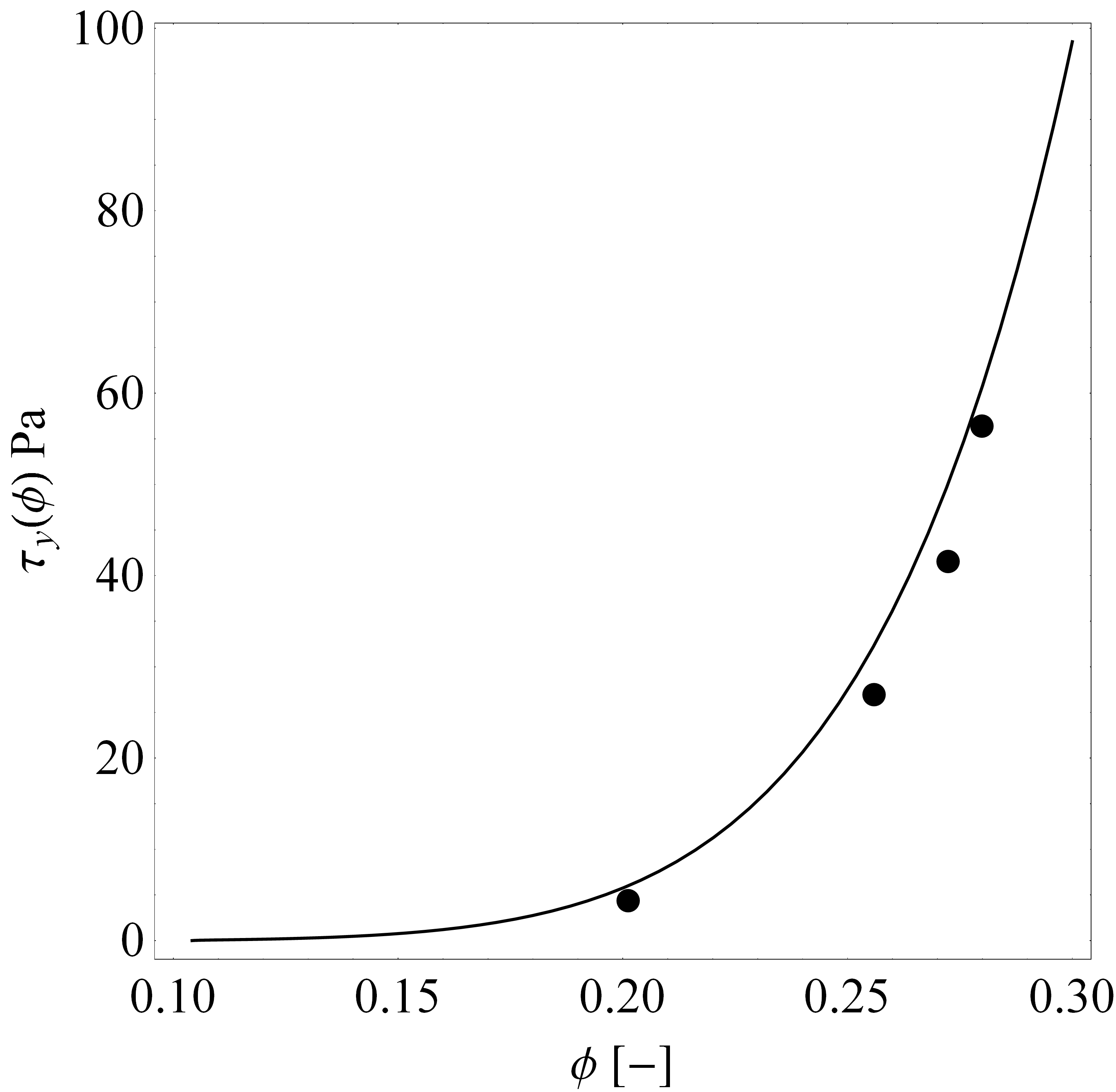}&
\includegraphics[width=0.32\columnwidth]{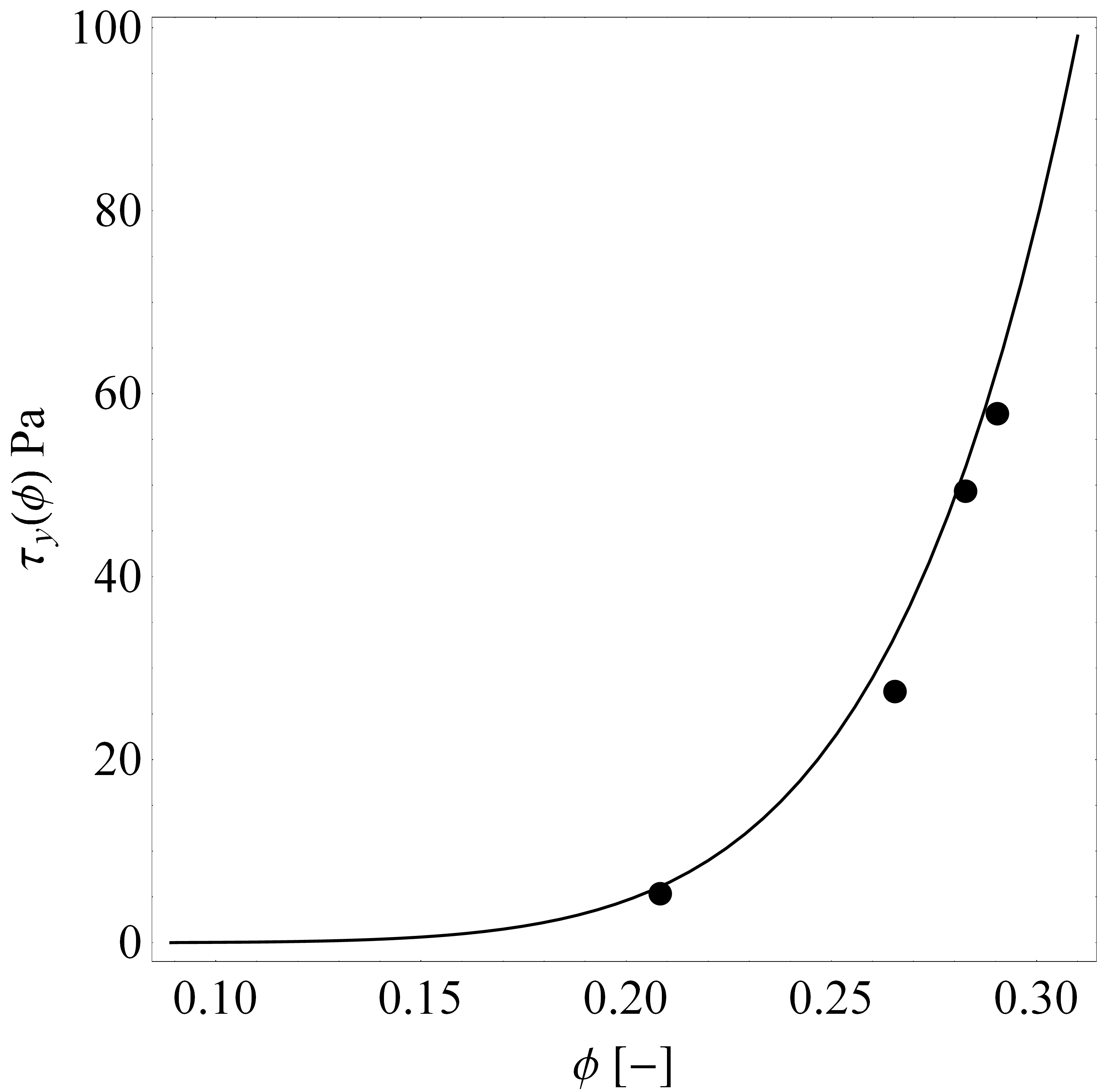}\\
(a) & (b) & (c)
\end{tabular}
\end{centering}
\caption{\emph{In-situ} measurements (points) and model predictions (lines) of the shear yield stress $\tau_y(\phi)$ for calcium carbonate suspensions under flocculant types and dosages (a)-(c) summarized in Table~\ref{tab:suspensions}.}\label{fig:shear_yield_stress}
\end{figure}

To validate the fits predicted from the viscoplastic model, \emph{in-situ} shear yield stress measurements were performed in the 110 mm column as per the protocol shown in Fig.~\ref{fig:vane_schematic}, using a Haake VT500 rheometer fitted with a cruciform vane (diameter 22 mm, height 31 mm) following the procedure of \citet{NguyenBoger:83}. As the shear vane spans a significant bed height (31 mm), the measured shear yield strength is an average of the shear yield strength distribution (arising from the solids volume fraction distribution) over the height of the vane. These average shear yield strengths are shown in Table~\ref{tab:rheometry}, and the corresponding average solids volume fraction for each measurement is determined from the fitted solids volume fraction profile for the 110 mm column shown in Fig.~\ref{fig:profiles_fitted}. These averaged values are shown in Fig.~\ref{fig:shear_yield_stress} as data points, and the shear yield stress functions predicted from fitting of the profile data in Fig.~\ref{fig:profiles_fitted} are shown as continuous curves. The fitted shear yield stress functions agree within experimental error (estimated to be order 20\%) of the \emph{in-situ} vane rheometer measurements. Also note that the accuracy of the fitted shear yield stress function is contingent upon both the validity of the assumption $\tau_y(\phi)\approx\tau_w(\phi)$, and linearisation of $\tau_y(\phi)$ over the vane height during the averaging process.

\begin{figure}
\begin{centering}
\begin{tabular}{c c}
\includegraphics[width=0.45\columnwidth]{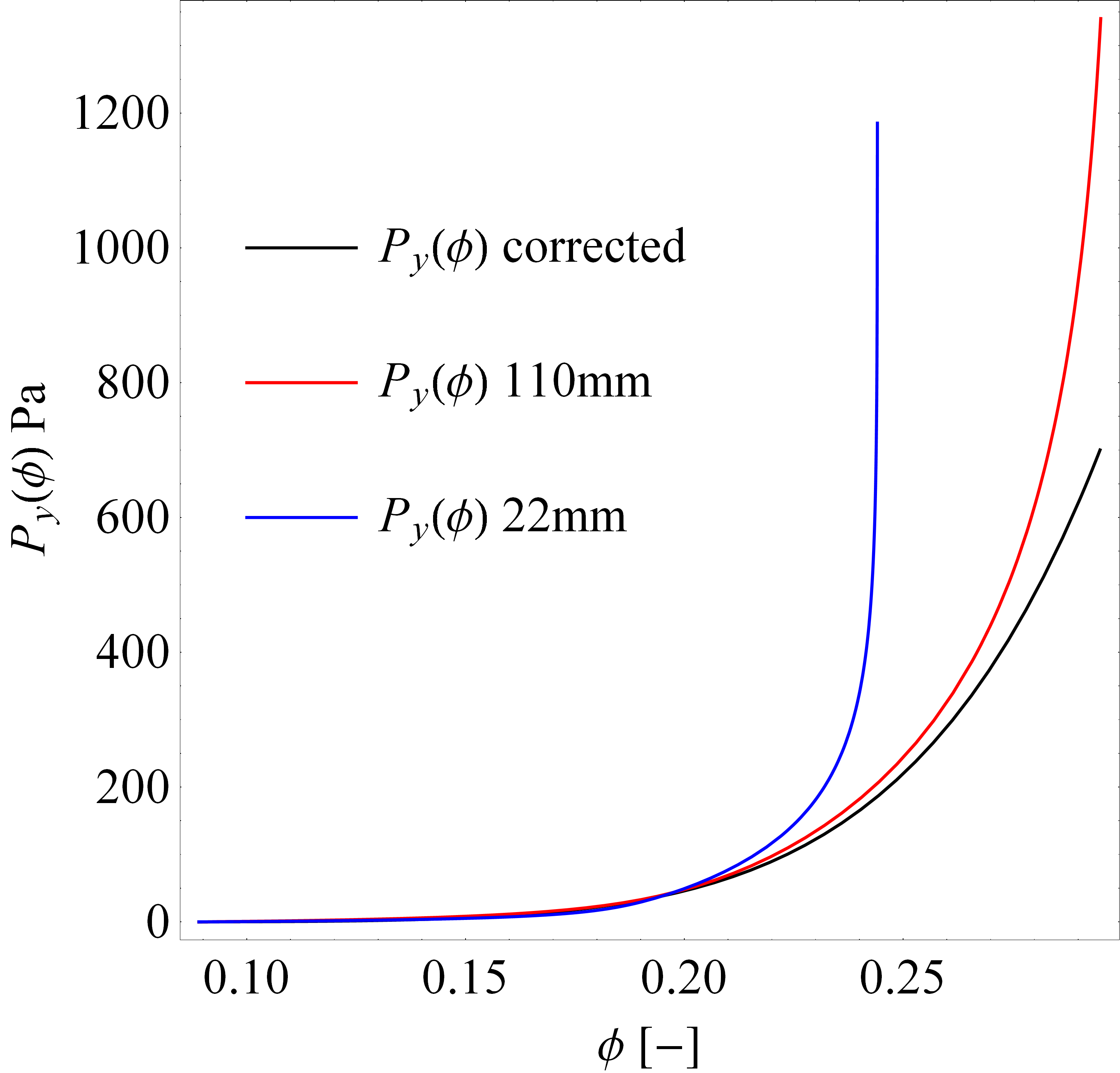}&
\includegraphics[width=0.45\columnwidth]{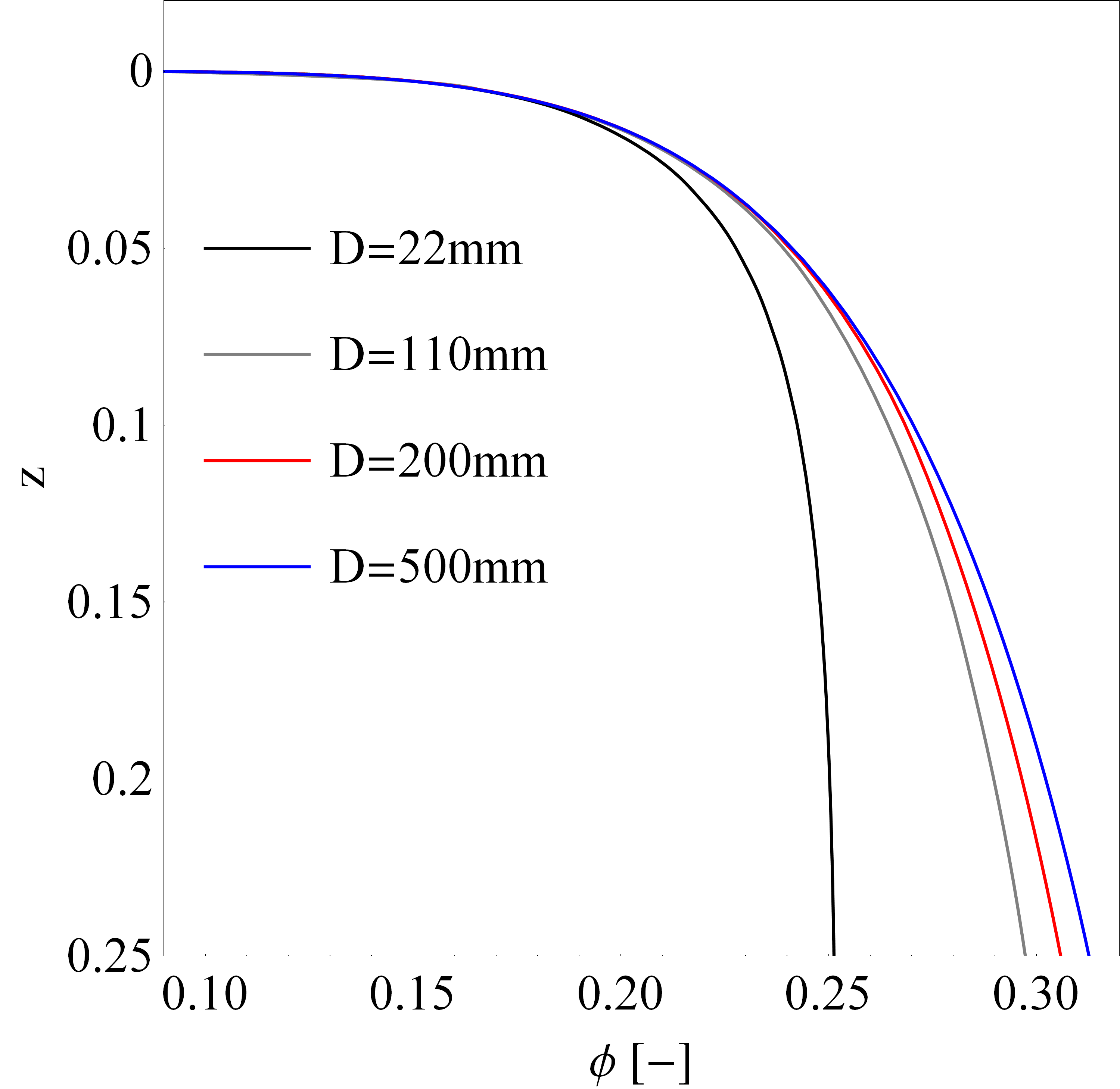}\\
(a) & (b)
\end{tabular}
\end{centering}
\caption{(a) Errors for uncorrected (i.e. not accounting for wall adhesion) estimation of the compressive yield stress $P_y(\phi)$ for suspension (c) in Table~\ref{tab:suspensions} in the 22 mm and 110 mm columns, and (b) prediction of equilibrium solids volume fraction profiles for suspension (c) in Table~\ref{tab:suspensions} across various column diameters.}\label{fig:py_error}
\end{figure}

Agreement between the viscoplastic model and \emph{in-situ} shear yield strength and volume fraction profile measurements suggests that the ratio $S(\phi)$ of shear to compressive yield strength for these colloidal gels is significantly higher (\~0.1-1) than previously reported, and as such wall adhesion effects are also much more prevalent. Neglect of wall adhesion effects can lead to very serious errors in the estimation of the compressive yield strength, as illustrated in Fig.~\ref{fig:py_error}(a), which shows significant deviations from the corrected compressive yield strength in both the 22 mm and 110 mm columns, where the divergence of $P_y(\phi)$ for the uncorrected estimate in the 22 mm column around $\phi\approx 0.23$ arises from the vertical solids volume fraction profile in Fig.~\ref{fig:profiles}. Fig.~\ref{fig:py_error}(a) also suggests significant wall effects arise in the 110 mm column, where the uncorrected compressive yield strength estimate deviates by up to 100\% at large $\phi$. Traditionally, such wide columns would have been considered free from wall adhesion effects, and even very wide columns still exhibit significant errors as shown in Fig.~\ref{fig:py_error}(b).

For such batch sedimentation problems involving significant wall adhesion effects, the viscoplastic formulation (\ref{eqn:omega1}) (\ref{eqn:omega2}) under the closure approximation $N_1=N_2=0$, and wall adhesion assumption $\tau_y(\phi)\approx\tau_w(\phi)$ serves as a useful analysis tool, and leads to a useful methodology to generate accurate estimates of the shear $\tau_y(\phi)$ and compressive $P_y(\phi)$ yield strengths from the solids volume fraction profile data across several columns of various widths.

\begin{centering}
\begin{figure}
\centering
\includegraphics[width=0.45\columnwidth]{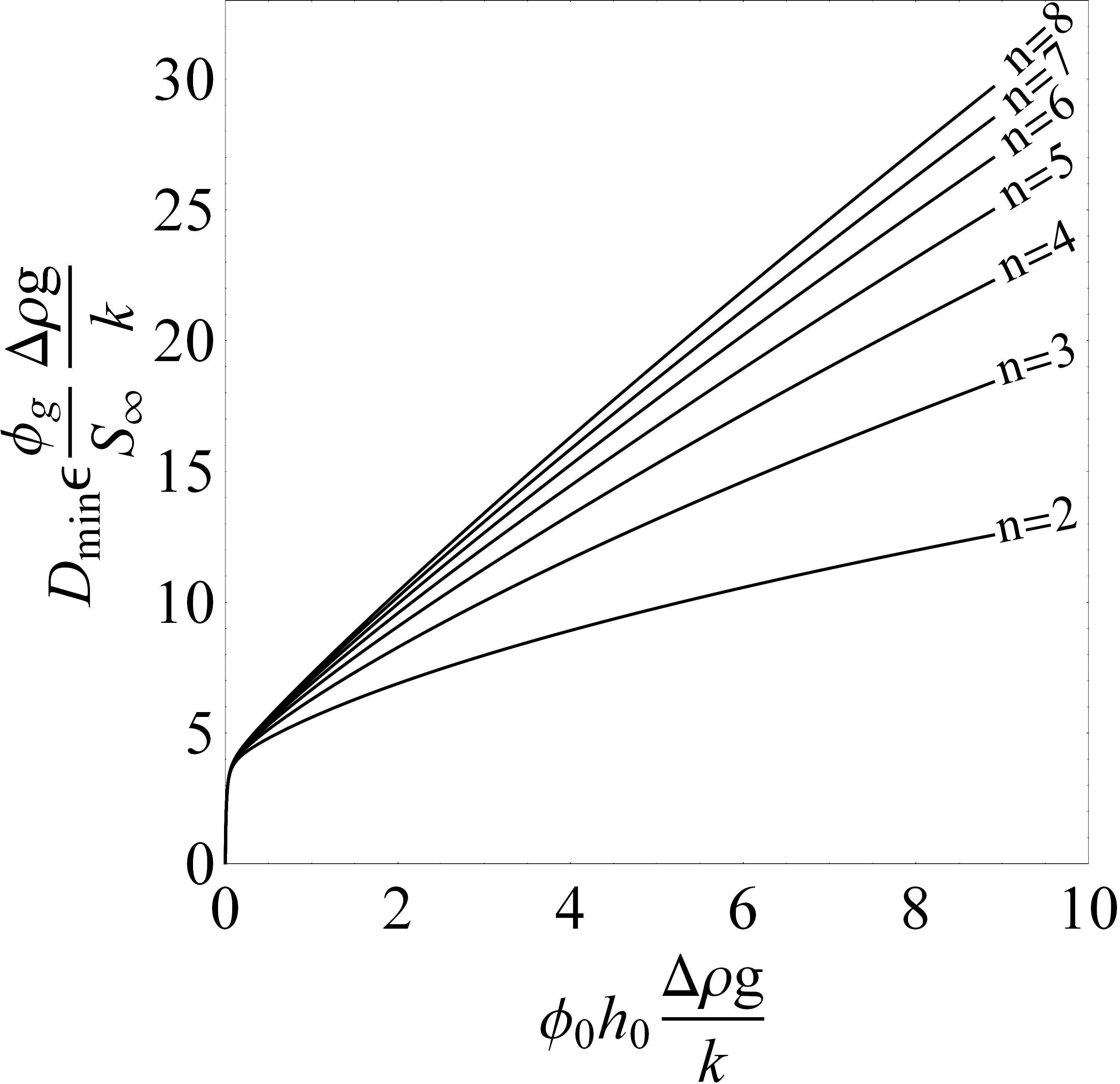}
\caption{Scaled minimum column diameter $D_{\text{min}}$ and relative error $\epsilon$ as a function of scaled equilibrium network pressure $p_\infty/k$ for various indices $n$ for functional forms (\ref{eqn:pyphi_func}), (\ref{eqn:Sphi}).}\label{fig:wall_error}
\end{figure}
\end{centering}

A useful experimental parameter is the minimum column diameter $D_{\min}$ required to render wall adhesion effects negligible. To derive a relationship between $D_{\min}$ and the relevant experimental and suspension parameters, we define the relative error $\epsilon$ from the vertical force balance (\ref{eqn:1Dbalance}) as the relative contribution of wall adhesion stress
\begin{equation}
\epsilon\equiv\frac{\frac{4}{D_{\min}}\tau_w(\phi)}{\Delta\rho g\phi},\label{eqn:epsilon}
\end{equation}
where in the limit of vanishing $\epsilon$, gravitational stress is balanced by the compressive yield strength of the suspension, and a 1D stress analysis is valid. Under the assumption that wall shear strength is well approximated by the bulk shear strength, $\tau_w(\phi)\approx \tau_y(\phi) = S(\phi)P_y(\phi)$, then for particulate gels whose shear yield strength is well characterized by (\ref{eqn:pyphi_func}) and (\ref{eqn:Sphi}), (\ref{eqn:epsilon}) may be expressed as
\begin{equation}
D_{\min}\epsilon\frac{\phi_g}{S_\infty}\frac{\Delta\rho g }{k}=\frac{4}{S_\infty}
\frac{\frac{p_\infty}{k}\left(1+\frac{p_\infty}{k}\right)^{1-\frac{1}{n}}}
{1+\frac{p_\infty}{k}\frac{1}{S_\infty}},\label{eqn:error}
\end{equation}
where $p_\infty=\Delta\rho g \phi_0 h_0$ is the equilibrium network pressure at the base of a batch settling experiment with initial height $h_0$ and volume fraction $\phi_0$. From (\ref{eqn:Sphi}), $S(\phi)$ decays rapidly from $S\sim 1$ toward the asymptotic value $S\rightarrow S_\infty$ as $\phi$ increases from $\phi_g$ for $n>2$, and so (\ref{eqn:error}) only varies very weakly with changes in $S_\infty$ for $S_\infty<0.1$. Under these physically reasonable conditions, the right hand side of (\ref{eqn:error}) only varies significantly with $p_\infty/k$ and the index $n$, and this relationship is plotted in Fig.~\ref{fig:wall_error} for various indices $n$ and relative stress $p_\infty/k$.

\begin{table}[b]
\begin{centering}
\begin{tabular}{c c c c c c c c}
\hline
Suspension & $p_\infty$ [Pa] & $S_\infty$ [-] & $D_{\text{min}}$ [mm]
& $z_{\text{max}, D=22 \text{mm}}$ [mm]
& $z_{\text{max}, D=110 \text{mm}}$ [mm]
& $\phi_{\text{max}, D=22 \text{mm}}$ [-]
& $\phi_{\text{max}, D=110 \text{mm}}$ [-]\\
\hline
(a) & 1142.9 & 0.157 & 3197 & 0.649 & 8.520 & 0.0966 & 0.1274 \\
(b) & 1125.1 & 0.112 & 2016 & 2.488 & 13.962 & 0.1436 & 0.1872 \\
(c) & 1137.6 & 0.113 & 1960 & 2.714 & 14.033 & 0.1496 & 0.1955 \\
\hline
\end{tabular}
\caption{Estimation of minimum column diameter $D_{\text{min}}$, maximum bed depth $z_{\text{max}}$, and maximum volume fraction $\phi_{\text{max}}$ for suspensions (a)-(c) in 22 mm and 110 mm diameter columns for an experimental error $\epsilon$=5\%.}\label{tab:errors}
\end{centering}
\end{table}

Application of (\ref{eqn:error}) to the three suspensions considered experimentally clearly illustrates the prevalence of wall adhesion effects for high molecular weight polymer flocculated colloidal gels. For a modest experimental error $\epsilon=5$\%, minimum column diameters $D_{\min}$ of at least 2 m are required to render wall adhesion effects negligible. Conversely, one may calculate the maximum bed depth $z_{\max}$ and solids volume fraction $\phi_{\max}$ below which the error exceeds $\epsilon$. All of these measures indicate that either very shallow bed depths or impractically wide columns are required to avoid significant errors, hence correction of wall adhesion effects is necessary for many experiments involving strongly flocculated colloidal suspensions.

Development of this model for batch sedimentation with wall adhesion effects raises fundamental issues regarding the constitutive modeling of strongly flocculated colloidal gels, specifically the validity and utility of the viscoplastic and hyperelastic formulations. Whilst selection of an appropriate modelling framework is contingent upon the application at hand and the requisite level of fidelity and tractability, an understanding of the nature and range of validity of each approach is of critical importance. It is anticipated that such issues shall play a central role in the continuing development and application of constitutive models for strongly flocculated colloidal gels.

\section{Conclusions}
\label{sec:conclusions}
The assumption that wall adhesion effects are negligible for the batch settling of strongly flocculated colloidal gels is commonly invoked via the justification that the shear yield strength is small compared to the compressive yield strength. In this study, \emph{in-situ} measurements of both colloidal gel rheology and solids volume fraction distribution in equilibrium batch settling experiments suggest the contrary for polymer flocculated colloidal gels, where wall adhesion effects are found to be significant in a 110 mm diameter column, normally considered to be sufficient to render wall effects negligible. Neglect of such effects can lead to serious errors in estimation of e.g. the compressive yield stress, where errors of order 100\% are observed at higher concentrations in the 110 mm diameter column, and divergence of the compressive yield stress for a 22 mm diameter column.

Consideration of a mathematical model for the batch settling equilibrium stress state in the presence of wall adhesion raises fundamental issues regarding the constitutive modeling of strongly flocculated colloidal gels, namely the relative utility of viscoplastic and viscoelastic rheological models under arbitrary tensorial loadings. More commonly used viscoplastic models (e.g. generalisation of Herschel-Bulkely or Bingham models to the compressible case~\citep{LesterEA:10,Stickland:09,Michaels/Bolger:62}) quantify the shear and compressive rheology of colloidal gels solely in terms of critical yield strength, ignoring the detailed mechanisms of shear strain softening and compressive strain hardening, whereas hyperelastic models treat the particulate network as a history-dependent viscoelastic material which facilitates detailed resolution of the complex rheological behaviour~\citep{LindstomEA:12,SprakelEA:11,GibaudEA:10,Santos:13,Koumakis:11,GibaudEA:08,Ovarlez:13,RamosEA:01,Ovarlez:07,CloitreEA:00,Tindley:07,Kumar:12,Uhlherr:05,GrenardEA:14} inherent to colloidal gels.

In the context of batch settling with wall adhesion effects, the viscoplastic formulation leads to a statically indeterminate formulation due to the strain being undefined for stresses below the yield value. This formulation is closed by assuming the first normal stress difference is negligible, and this closure is found to be a reasonable approximation by conversion of the viscoplastic solution to the hyperelastic frame and evolution to the hyperelastic equilibrium state. Whilst the viscoplastic model is appropriate for this problem, the hyperelastic formulation is required to resolve a wide class of colloidal gel flow phenomena, and it is of critical importance to determine the limitations of the viscoplastic model in a given application.

Application of the viscoplastic model to the \emph{in situ} measurements serves as an indirect validation of this model, and points to a methodology for estimation of the shear and compressive yield strengths from a series of batch settling experiments in various width columns. These estimates are shown to fall within experimental error of the \emph{in-situ} shear yield stress measurements, and furthermore suggest the strength ratio and hence wall adhesion effects in strongly flocculated colloidal gels can be much greater than was appreciated hitherto. They should always be expected near the gel-point where the ratio of shear to compressive strength is largest and it has been found that they can be much stronger overall for particles flocculated with high molecular weight polymers than has been found for coagulated systems.


\section{Acknowledgements}
\label{sec:acknow}
This work was conducted as part of AMIRA P266F ``Improving Thickener Technology'' project, supported by the following companies: Alcoa World Alumina, Alunorte, Anglo American, BASF, Bateman Engineering, BHP Billiton, Cytec Australia Holdings, Exxaro, FL Smidth Minerals, Freeport-McMoran, Hatch, Metso, MMG, Nalco, Outotec, Rio Tinto, Rusal, Shell Energy Canada, Teck Resources, Total E\&P Canada and WesTech. The authors are indebted to Jon Halewood for undertaking the flocculation, rheological and solid volume fraction profile measurements, Andrew Chryss for design of rheological measurements, and Michel Tanguay for valuable discussions.


\end{document}